\documentclass[10pt,letterpaper]{aastex6}
\usepackage{epsfig}

\def\fluxthres{\hat f_{\bar \e}}

\def\zmax{z_{\rm max}}

\def\e{\epsilon}

\def\Estarg{{\cal E}_{*\gamma}}

\def\Swift{\emph{Swift}}
\def\Fermi{\emph{Fermi}}

\newcommand{\begeq}{\begin{equation}}
\newcommand{\fineq}{\end{equation}}
\newcommand{\begfig}{\begin{figure}}
\newcommand{\finfig}{\end{figure}}
\newcommand{\begeqarray}{\begin{eqnarray}}
\newcommand{\fineqarray}{\end{eqnarray}}

\slugcomment{Accepted for publication to MNRAS}

\shorttitle{Excess of Long GRB's at Low Redshift} 

\shortauthors{Le, Ratke \& Mehta}

\begin{document}

\title{Resolving the Excess of Long GRB's at Low Redshift in the Swift era}

\author{Truong Le$^{1,2}$, Cecilia Ratke$^{1}$, and Vedant Mehta$^{1,3}$}

\affil{$^{1}$Department of Physics, Astronomy \& Geology, Berry College, Mount Berry, GA 30149, USA; tle@berry.edu;
\\$^{2}$Department of Physics, American University, Washington, DC 20016, USA; tle@american.edu;
\\$^{3}$Now at: Los Alamos National Laboratory, P.O. Box 1663, Los Alamos, NM 87545, USA}

\begin{abstract}
Utilizing more than 100 long gamma-ray bursts (LGRBs) in the \Swift-Ryan-2012 sample that includes the observed redshifts and jet angles, Le \& Mehta performed a timely study of the rate density of LGRBs with an assumed broken power-law GRB spectrum and obtained a GRB-burst-rate functional form that gives acceptable fits to the pre\Swift~and \Swift~redshift, and jet angle distributions. The results indicated an excess of LGRBs at redshift below $z \sim 2$ in the \Swift~sample. In this work, we are investigating if the excess is caused by the cosmological Hubble constant $H_0$, the gamma-ray energy released $\Estarg$, the low- and high-energy indices ($\alpha, \beta$) of the Band function, the minimum and maximum jet angles $\theta_{\rm j,min}$ and $\theta_{\rm j, max}$, or that the excess is due to a bias in the \Swift-Ryan-2012 sample. Our analyses indicate that none of the above physical parameters resolved the excess problem, but suggesting that the \Swift-Ryan-2012~sample is biased with possible afterglow selection effect. The following model physical parameter values provide the best fit to the \Swift-Ryan-2012 and pre\Swift~samples: the Hubble constant $H_0 = 72 \, {\rm km s^{-1} Mpc^{-1}}$, the energy released $\Estarg \sim 4.47 \times 10^{51}$ erg,  the energy indices $\alpha \sim 0.9$ and $\beta \sim -2.13$, the jet angles of $\theta_{\rm j,max} \sim 0.8$ rad, and $\theta_{\rm j, min} \sim 0.065$ and $\sim 0.04$ rad for pre\Swift~and \Swift, respectively, $s \sim -1.55$ the jet angle power-law index, and a GRB formation rate that is similar to the Hopkins \& Beacom observed star formation history and as extended by Li. Using the \Swift~Gamma-Ray Burst Host Galaxy Legacy Survey (SHOALS) \Swift-Perley LGRB sample and applying the same physical parameter values as above, however, our model provides consistent results with this data set and indicating no excess of LGRBs at any redshift.
\end{abstract}


\section{INTRODUCTION}
With the launch of the \Swift~satellite, rapid follow-up studies of gamma ray bursts (GRBs) triggered by the Burst Alert Telescope (BAT) on \Swift~became possible. Before 2008, the mean redshift of 41 pre\Swift~long-duration GRBs (LGRBs, which is the only type considered in this paper with $T_{90} > 2$ s) that also have measured beaming breaks is $\langle z\rangle = 1.5$~\citep[][hereafter pre\Swift-Friedman sample]{fb05}, while 16 GRBs discovered by \Swift~have $\langle z\rangle = 2.72$~\citep[][hereafter \Swift-Jakobsson sample]{jak06}. The mean measured redshifts of the LGRBs indicate that a fainter and more distant population of GRBs than found with the pre\Swift~satellites BATSE, BeppoSAX, INTEGRAL, and HETE-2 was detected. However, it is important to notice that there were many GRBs detected by the pre\Swift~instruments that did not have measured redshifts~\citep[e.g.,][]{sn19}, particularly BATSE, with more than 2700 detected GRBs~\citep[e.g.,][]{pac04}. Hence, it is very likely that BATSE also observed a population of faint and distant GRBs as well, since BATSE's effective area was large enough that it was more sensitive to high- z GRBs than most current instruments~\citep[e.g.,][]{tas08}. From these initial samples, \citet{bag06} performed statistical tests to compare the redshift distributions of \Swift~and pre\Swift~GRBs and concluded that the redshift distributions are different~\citep[also see Figure~3 from][]{jak06}, suggesting that the pre\Swift~and \Swift~data sample distinct subsets of the overall GRB population due to different detector characteristics and follow-up capabilities. The pre\Swift-Friedman and \Swift-Jakobsson LGRB samples are presented in Table~\ref{tbl-1}, where the pre\Swift~sample contains both the redshift and jet opening angle values, while the \Swift~sample only contains the redshift values because there are problems with estimating the achromatic jet break in the Swift data \citep[e.g.,][]{kz15}.
%
\begin{deluxetable}{|lcccc||lc|}
\tabletypesize{\small} %
\tablecaption{{\bf The Friedman and the Jakobsson samples}
\label{tbl-1}} %
\tablewidth{-0.pt} %
\tablehead{ %
\colhead{$\rm GRB^a$} %
&\colhead{$\rm z$}%
&\colhead{$\rm \theta_{jet} (deg)$} %
&\colhead{$\rm \alpha$} %
&\colhead{$\rm \beta$ }  %
&\colhead{$\rm GRB^b$} %
&\colhead{$\rm z$}%
} \startdata
970508
& 0.8349
& 21.83
& -1.71				
& -2.20
& 050315
&1.95
\\
970828
& 0.9578
& 7.26
& -0.70				
& -2.07
& 050318
& 1.44
\\
971214        
& 3.418 
& 5.48        
& -0.76				  
& -2.70  
& 050319
& 3.24
\\ 
980326 
& 1
& 6.33
& -1.23				
& -2.48
& 050401
& 2.90
\\
980519
& 2.5
& 3.65
& -1.35				
& -2.30
& 050416A
& 0.65
\\
980613
& 1.0969								
& 12.82
& -1.43				
& -2.70
& 050505
& 4.27
\\
980703 
& 0.9662				
& 11.42
& -1.31				
& -2.40
& 050525
& 0.61
\\
981226
& 1.5				
& 14.80
& -1.25				
& -2.60
& 050603
& 2.82
\\
990123
&1.6004				
& 4.68
& -0.89				
& -2.45
& 050730
& 3.97
\\
990510 
& 1.6187				
& 3.77
& -1.23				
& -2.70
& 050802
& 1.71
\\
990705
& 0.8424				
& 5.5
& -1.05				
& -2.20
& 050814
& 5.3
\\
990712
& 0.4331				
& 11.78
& -1.88				
& -2.48
& 050820A
& 2.61
\\
991208
& 0.7055				
& 8.39
& --
& --
& 050824
& 0.83
\\
991216
& 1.0200				
& 4.66
& -1.23				
& -2.18
& 050904
& 6.29
\\
000131
& 4.5				
& 4.71
& -1.20				
& -2.40
& 050908
& 3.34
\\
000210
& 0.8463				
& 5.43
& --
& --
& 050922B
& 2.20
\\ \cline{6-7}
000301C
& 2.0335				
& 13.88
& --
& --
\\
000418
&1.1182				
&22.95
& --
& --
\\
000630
& 1.5				
& 9.85
& --
& --
\\
000911
& 1.0585				
& 5.58
& -1.11				
& -2.32
\\
000926
& 2.0369				
& 6.28
& --
&--
\\
010222
& 1.4769				
& 3.29
& -1.35				
& -1.64
\\ 
010921
& 0.4509				
& 32.76
& -1.55				
& -2.30
\\
011121
& 0.3620				
& 16.24
& -1.42				
& -2.30
\\
011211
& 2.14				
& 5.98
& -0.84				
& -2.30
\\
020124
& 3.198				
& 11.3
& -0.79				
& -2.30
\\
020405
& 0.6899				
& 7.68
& --
& --
\\
020427
& 2.30				
&18.52
& -1.00				
& -2.10
\\
020813
& 1.254				
& 3.24
& -0.94				
& -1.57
\\
021004
& 2.3351				
& 12.73
& -1.01				
& -2.30
\\
021211
& 1.0060				
& 8.78
& -0.86				
& -2.18
\\
030226 
&1.9860				
& 4.99
& -0.89				
& -2.30 
\\
030323
& 3.3718				
& 5.71
& -1.62				
& -2.30
\\
030328
&1.52				
& 4.37
& -1.14				
& -2.09
\\ 
030329
& 0.1685				
& 6.6
& -1.26				
& -2.28  
\\
030429
& 2.6564				
& 7.41
& -1.12				
& -2.30
\\
030528 
&1				
& 9.27
& -1.33				
& -2.65
\\
030723 
& 2.1				
& 11.89
& -1.00				
& -1.90
\\
040511
& 1.5				
& 5.91
& -0.67				
& -2.30
\\
040924
& 0.8590				
& 8.36
& -1.17				
& -2.30
\\
041006
& 0.7160				
& 8.11
& -1.37				
& -2.30
\\
\enddata
\tablecomments{$^a$These are the pre\Swift~data from the \citet{fb05}.
$^b$These are the \Swift~data from the \citet{jak06}.}
\end{deluxetable}

\citet[][hereafter LD07]{ld07} developed a physical model to understand the differences between the redshift and jet opening angle distributions of the initial samples \citep[see Figure~2 of][]{ld07}, taking into account the different detector triggering properties. Their GRB model is based on the uniform jet model and a flat $\nu F_\nu$ GRB spectrum with an assumption that the LGRB density rate is proportional to the measured star formation rate (SFR), where the SFR7 in the inset of Figure~\ref{fig1}(a) is the observed \citet{hb06} SFR history. LD07 assumed a flat $\nu F_\nu$ GRB spectrum as their initial study to understand the observed LGRB redshift distribution, but a more realistic model is a broken power-law $\nu F_\nu$ GRB spectrum, which was later studied by \citet{lm17} as discussed below. LD07, however, showed that a good fit was only possible by providing a positive evolution of the SFR history of GRBs to high redshifts that is similar to SFR5 or SFR6, see the inset of Figure~\ref{fig1}(a).  These SFR analytic fitting profiles are given by $  \Sigma_{_{\rm SFR}}(z) = \frac{1+(\eta_{_2} z/\eta_{_1})}{1+(z/\eta_{_3})^{\eta_{_4})}}$ as a function of redshift $z$ with the parameter values of $\eta_{_1}, \eta_{_2}, \eta_{_3}$, and $\eta_{_4}$ for SFR5, SFR6, SFR7 as given in Table~\ref{tbl-2}. The first, second, and last columns of Table~\ref{tbl-2} are the GRB spectral model type, the GRB density rate model, and the reference of the paper that assumes these models and values, respectively. The third row represents the SFR analytic fitting profile for the observed SFR that was obtained by \citet{hb06}.  LD07 calculated LGRB density rate models, SFR5 and SFR6, at high redshift that are consistent with other researchers \citep[e.g.,][]{lfr02,dai06,mes06,gp07,kis08,wp10}. However, it remains unclear whether the excess at high redshift is due to luminosity evolution \citep[e.g.,][]{sc07,sal09,sal12,den16} or the cosmic evolution of the GRB rate \citep[e.g.,][]{but10,qin10,wp10}. Furthermore, \citet{sal12}, for example, suggested that a broken power-law luminosity evolution with redshift is required to fit the observed redshift distribution. 

Recently, \citet[][hereafter LM17]{lm17} revisited the work done by LD07 and performed a timely study of the rate density of GRBs with an assumed broken power-law GRB spectrum with the low- and high-energy indices $(\alpha=-1, \beta=-2.5)$ of the Band function \citep[][]{ban93}. Utilizing more than 100 LGRBs in the \Swift~sample that includes both the observed estimated redshifts and jet opening angles \citep[][hereafter \Swift-Ryan 2012 sample]{zha15,rya15}, LM17 obtained a GRB burst rate functional form (SFR9 - a long thin dashed line in Figure~\ref{fig1}(a)) that produces the model distributions that are comparable to the observed pre\Swift~and \Swift-Ryan 2012 redshift and jet opening angle distributions (see Figures~\ref{fig1}(b) and the inset of \ref{fig1}(b)). They also showed that the SFR5 and SFR6 models that were utilized by LD07 could not fit the more current updated LGRB \Swift-Ryan 2012 sample. The SFR9 analytic fitting profile is given by equation (18) in  LM17 (and in eq.~\ref{eq1} as discussed in Section~2) and their parameter values are also given in Table~\ref{tbl-2} with the breaking redshift at $z_{_1}$ and $z_{_2}$ equal to $0.5$ and $4.5$, respectively. Interestingly, SFR9 is similar to the \citet{hb06} star formation history (SFR7) and as extended by \citet{li08}. Most importantly, the result indicates an excess of LGRBs at low redshift below $z \sim 2$ in the \Swift~sample (see Figure~\ref{fig1}(b)), consistent with other researchers  \citep[e.g.][]{pkk15,yu15,laj19}. However, the reason for this excess is either unclear, incomplete sample size, or that GRB formation rate does not trace SFR at low redshift less than $z \leq 1$\citep[e.g.][]{pkk15,yu15,laj19}, and we plan to address these concerns in this work.  The \Swift-Ryan LGRB sample is presented in Table~\ref{tbl-3}.

%
\begin{deluxetable}{llccccc}
\tabletypesize{\small} %
\tablecaption{{\bf GRB formation rate model parameters}
\label{tbl-2}} %
\tablewidth{-0.pt} %
\tablehead{ %
\colhead{$\rm  $} %
&\colhead{$\rm GRB_{DRM}$}%
&\colhead{$\rm \eta_{_1}$}%
&\colhead{$\rm \eta_{_2}$} %
&\colhead{$\rm \eta_{_3}$} %
&\colhead{$\rm \eta_{_4}$} %
&\colhead{$\rm Ref.$} 
} \startdata
Flat $\nu F_{\nu}$ 
& SFR5
& 0.015
& 0.12
& 3.0
& 1.3
& LD07
\\
{ }
&SFR6
& 0.011
& 0.12
& 3.0
& 0.5
& LD07
\\
\hline
Observed star formation rate 
&SFR7$^a$
& 0.0157
& 0.118
& 3.23
& 4.66
& LD07, LM17, this paper	
\\
\hline
{ }
&SFR9        
& 4.1
& 0.8        
& -5.1
&  --  
& LM17   
\\ 
Broken power-law $\nu F_{\nu}$  
&SFR10
& 8
& -0.4
& -5.1
& --	
& this paper
\\
{}
&SFR11
& 5.5
& 0.38
& -4.1
& --
& this paper	
\\
\hline \hline
\enddata
\tablecomments{SFR5, SFR6, SFR9, SFR10, and SFR11 are the calculated GRB density rate model (DRM). Also, the dash line in the table just means the current model does not have the $\eta_{_4}$ parameter. The last column indicates which paper utilizes the corresponding model. $^a$The parameters values in the SFR7 model are the fitted values to the observed \citet{hb06} SFR history that were discussed in the LD07 paper.}
\end{deluxetable}
\begfig[t] \hskip-0.3in \vskip-0.15in \epsscale{1.15} \plottwo{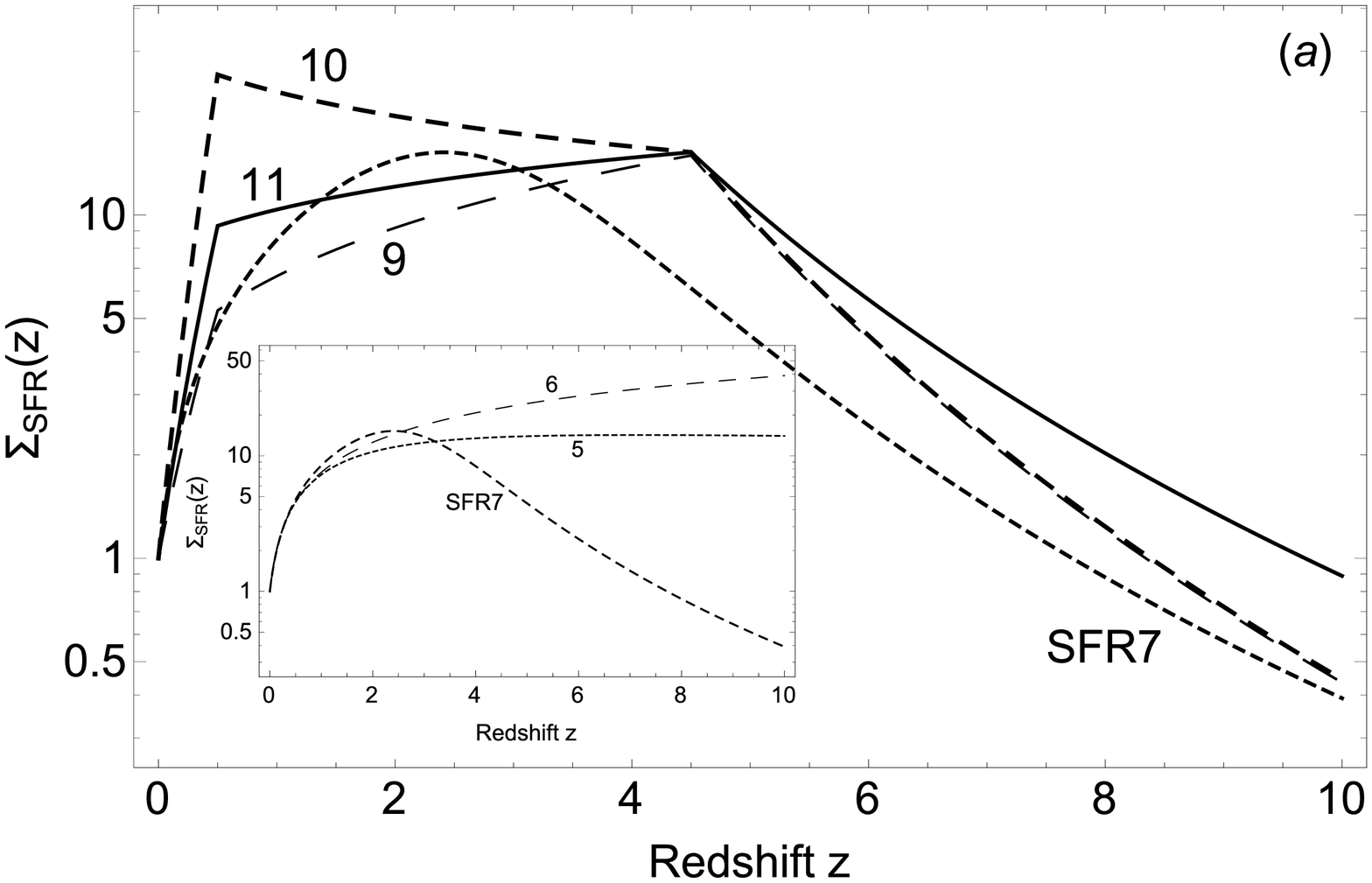}{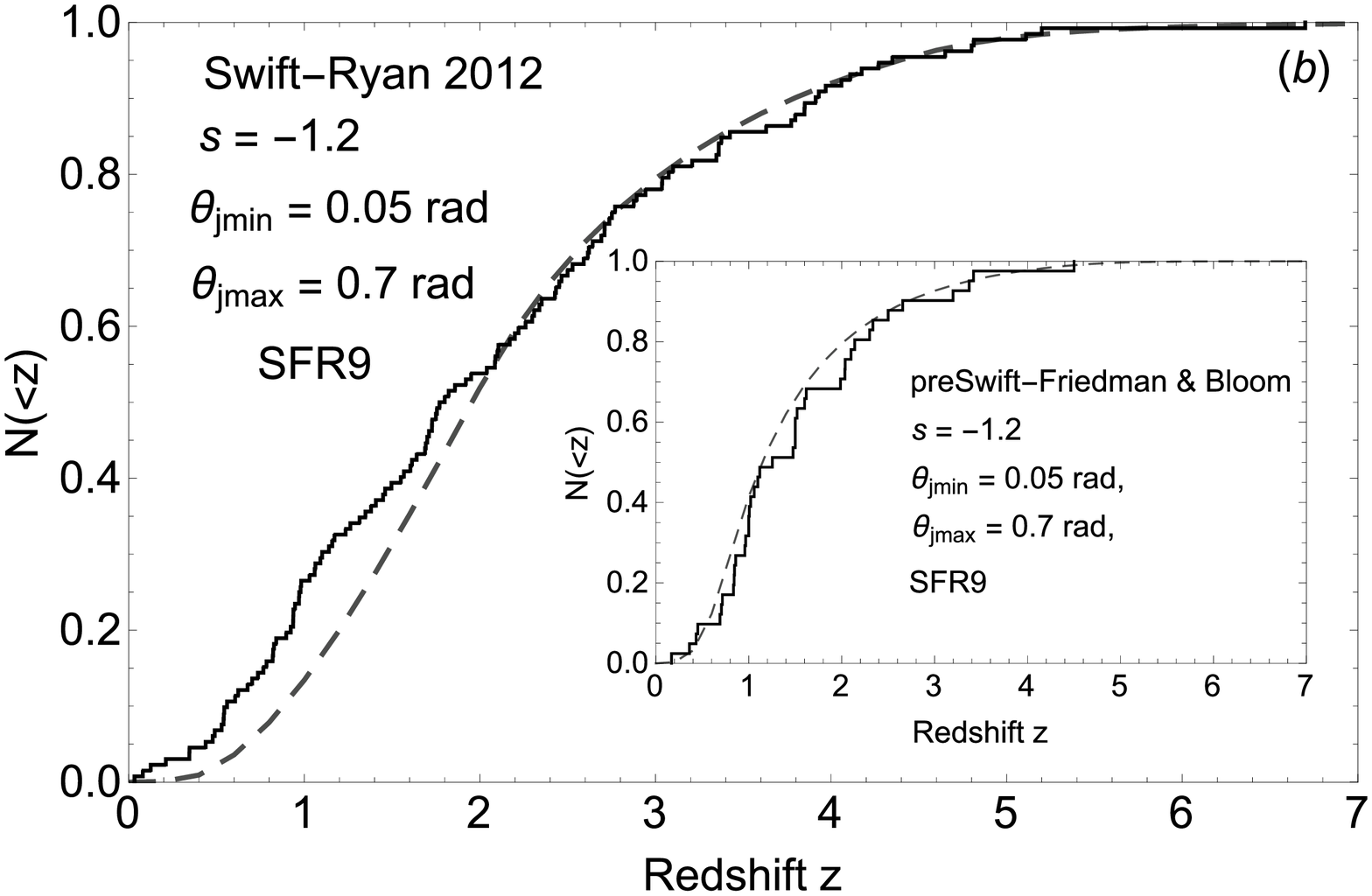}
\caption{\footnotesize (a) Curves 5 (SFR5) and 6 (SFR6) from the inset, and 9 (SFR9) are the GRB density rates that give an acceptable fit to the \Swift~and pre\Swift~redshift distribution assuming a GRB flat and a broken power-law spectrum, respectively,~\citep{lm17}. SFR7 from the inset is the~\citet{hb06} SFR history and as extended by~\citet{li08}. Curves 10 (SFR10) and 11 (SFR11) give an acceptable fit to the \Swift~and pre\Swift~redshift distribution, but SFR11 gives a better fit in this work. (b) The cumulative redshift distributions of 133 GRBs in the \Swift-Ryan-2012 sample~\citep[thick solid line;][]{rya15}, and the pre\Swift~sample~\citep[from the inset with thick solid line;][]{fb05}; the dashed lines are the calculated distributions. The inset of Figure 1(a) and Figures 1(b) are from \citet{ld07} and \citet{lm17} papers, respectively.}
\label{fig1} 
\finfig
%
\begin{deluxetable}{|lcc|lcc|lcc|}
\tabletypesize{\small} %
\tablecaption{{\bf \Swift-Ryan sample}
\label{tbl-3}} %
\tablewidth{-0.pt} %
\tablehead{ %
\colhead{$\rm GRB$} %
&\colhead{$\rm z$}%
&\colhead{$\rm \theta_{jet} (rad)$} %
&\colhead{$\rm GRB$} %
&\colhead{$\rm z$}%
&\colhead{$\rm \theta_{jet} (rad)$} %
&\colhead{$\rm GRB$} %
&\colhead{$\rm z$}%
&\colhead{$\rm \theta_{jet} (rad)$} %
} \startdata
121024A$^a$	
& 2.298	
& 0.0565$^{+0.0179}_{-0.008}$
&090328A	$^b$
&0.736	
&0.32$^{+0.13}_{-0.17}$	
&061007$^b$	
&1.2622	
&0.31$^{+0.13}_{-0.17}$	
\\
120922A$^b$	
& 3.1	
&0.3$^{+0.14}_{-0.13}$	
&090313$^b$	
&3.375	
&0.127$^{+0.143}_{-0.04}$	
&061006$^b$	
&0.4377	
&0.407$^{+0.068}_{-0.173}$	
\\
120909A$^b$	
&3.93	
&0.22$^{+0.18}_{-0.13}$	
&090205$^a$	
&4.6497	
&0.0513$^{+0.0081}_{-0.0046}$	
&060926$^b$	
&3.2086	
&0.29$^{+0.15}_{-0.16}$	
\\
120802A$^b$	
&3.796	
&0.21$^{+0.18}_{-0.15}$	
&090113$^b$	
&1.7493	
&0.3$^{+0.13}_{-0.14}$	
&060912A$^b$	
&0.937	
&0.227$^{+0.152}_{-0.07}$	
\\
120712A$^b$	
&4.1745	
&0.301$^{+0.125}_{-0.099}$	
&081222$^b$	
&2.77	
&0.0844$^{+0.0081}_{-0.0120}$	
&060904B$^{a,b}$	
&0.7029	
&0.083$^{+0.048}_{-0.015}$	
\\
120404A$^b$	
&2.876	
&0.27$^{+0.15}_{-0.15}$	
&081221$^b$	
&2.26	
&0.34$^{+0.11}_{-0.093}$	
&060714$^a$	
&2.7108	
&0.0557$^{+0.0104}_{-0.0072}$	
\\
120119A$^b$	
&1.728	
&0.0510$^{+0.0481}_{-0.0046}$	
&081203A$^b$	
&2.1	
&0.121$^{+0.058}_{-0.063}$	
&060708$^a$	
&1.92	
&0.0597$^{+0.0115}_{-0.008}$	
\\
120118B$^b$	
&2.943	
&0.29$^{+0.14}_{-0.15}$	
&081109$^b$	
&0.9787	
&0.26$^{+0.16}_{-0.14}$	
&060707$^b$	
&3.424	
&0.22$^{+0.19}_{-0.16}$	
\\
111211A$^b$ 
&0.478	
&0.3$^{+0.14}_{-0.15}$	
&081029$^b$	
&3.8479	
&0.1615$^{+0.0055}_{-0.0061}$	
&060614$^b$	
&0.1257	
&0.293$^{+0.122}_{-0.085}$	
\\
111209A$^b$	
&0.677	
&0.34$^{+0.11}_{-0.13}$	
&081028$^b$	
&3.038	
&0.3$^{+0.13}_{-0.15}$	
&060607A$^{a,b}$	
&3.0749	
&0.374$^{+0.095}_{-0.072}$	
\\
111107A$^b$	
&2.893	
&0.28$^{+0.15}_{-0.15}$	
&081007$^b$	
&0.5295	
&0.159$^{+0.183}_{-0.066}$	
&060605$^b$	
&3.773	
&0.04614$^{+0.18939}_{-0.00096}$	
\\
110918A$^b$	
&0.982	
&0.35$^{+0.11}_{-0.17}$	
&080928$^b$	
&1.6919	
&0.25$^{+0.16}_{-0.17}$	
&060522$^b$	
&5.11	
&0.3$^{+0.13}_{-0.14}$	
\\
110818A$^b$	
&3.36	
&0.24$^{+0.18}_{-0.16}$	
&080913$^b$	
&6.7	
&0.359$^{+0.099}_{-0.125}$	
&060319$^b$	
&1.172	
&0.32$^{+0.12}_{-0.14}$	
\\
110808A$^b$	
&1.348	
&0.19$^{+0.19}_{-0.13}$	
&080810$^b$	
&3.3604	
&0.34$^{+0.11}_{-0.27}$	
&060218$^b$	
&0.0331	
&0.33$^{+0.12}_{-0.16}$	
\\
110801A$^{a,b}$	
&1.858	
&0.419$^{+0.056}_{-0.075}$	
&080721$^{a,b}$	
&2.5914	
&0.1117$^{+0.0109}_{-0.0083}$	
&060210$^a$	
&3.9122	
&0.0604$^{+0.0236}_{-0.0093}$	
\\
110715A$^b$	
&0.82	
&0.3$^{+0.14}_{-0.17}$	
&080607$^b$	
&3.0368	
&0.26$^{+0.16}_{-0.14}$	
&060206$^b$	
&4.059	
&0.377$^{+0.084}_{-0.111}$	
\\
110503A$^{a,b}$	
&1.613	
&0.1057$^{+0.0302}_{-0.0079}$	
&080605$^{a,b}$	
&1.6403	
&0.36$^{+0.087}_{-0.108}$	
&060202$^b$	
&0.785	
&0.28$^{+0.15}_{-0.15}$	
\\
110422$^{a,b}$	
&1.77	
&0.075$^{+0.025}_{-0.012}$	
&080603A$^b$	
&1.6880	
&0.31$^{+0.13}_{-0.16}$	
&060123$^b$	
&0.56	
&0.32$^{+0.13}_{-0.15}$	
\\
110213A$^b$	
&1.46	
&0.29$^{+0.14}_{-0.15}$	
&080430$^b$	
&0.767	
&0.0553$^{+0.0052}_{-0.0079}$	
&051111$^b$	
&1.55	
&0.28$^{+0.15}_{-0.15}$	
\\
110205A$^b$	
&2.22	
&0.388$^{+0.077}_{-0.145}$	
&080413B$^a$	
&1.1014	
&0.117$^{+0.02}_{-0.021}$	
&051109B$^b$	
&0.08	
&0.06$^{+0.162}_{-0.012}$	
\\
110128A$^b$	
&2.339	
&0.469$^{+0.023}_{-0.047}$	
&080413A$^b$	
&2.433	
&0.26$^{+0.16}_{-0.15}$	
&051016B$^b$	
&0.9364	
&0.35$^{+0.11}_{-0.24}$	
\\
100906A$^a$	
&1.727	
&0.054$^{+0.0038}_{-0.0079}$	
&080411$^b$	
&1.0301	
&0.097$^{+0.362}_{-0.017}$	
&051006$^b$	
&1.059	
&0.29$^{+0.15}_{-0.15}$	
\\
100901A$^a$	
&1.408	
&0.413$^{+0.033}_{-0.033}$	
&080319B$^{a,b}$	
&0.9382	
&0.098$^{+0.046}_{-0.013}$	
&050922C$^{a,b}$	
&2.1995	
&0.074$^{+0.033}_{-0.011}$	
\\
100728B$^b$	
&2.106	
&0.26$^{+0.15}_{-0.17}$	
&080310$^b$	
&2.4274	
&0.097$^{+0.263}_{-0.052}$	
&050915A	$^b$
&2.5273	
&0.24$^{+0.18}_{-0.14}$	
\\
100728A$^{a,b}$	
&1.567	
&0.112$^{+0.052}_{-0.011}$	
&080210$^b$	
&2.6419	
&0.164$^{+0.147}_{-0.098}$	
&050908$^b$	
&3.3467	
&0.26$^{+0.14}_{-0.13}$	
\\
100621A$^b$	
&0.542	
&0.0477$^{+0.0047}_{-0.002}$	
&080207$^b$	
&2.0858	
&0.31$^{+0.13}_{-0.14}$	
&050824$^b$	
&0.8278	
&0.140$^{+0.229}_{-0.088}$	
\\
100513A$^b$	
&4.8	
&0.28$^{+0.15}_{-0.15}$	
&080129$^b$	
&4.349	
&0.19$^{+0.2}_{-0.12}$	
&050820A$^{a,b}$	
&2.6147	
&0.151$^{+0.057}_{-0.022}$	
\\
100425A$^b$	
&1.755	
&0.16$^{+0.22}_{-0.10}$	
&071122$^b$	
&1.14	
&0.25$^{+0.16}_{-0.16}$	
&050801$^b$	
&1.38	
&0.29$^{+0.14}_{-0.12}$	
\\
100424A$^b$	
&2.465	
&0.17$^{+0.22}_{-0.11}$	
&071112C$^b$	
&0.8227	
&0.398$^{+0.071}_{-0.117}$	
&050730$^b$	
&3.9693	
&0.108$^{+0.301}_{-0.058}$	
\\
100418A$^b$	
&0.6235	
&0.23$^{+0.16}_{-0.12}$	
&071031$^b$	
&2.6918	
&0.3$^{+0.13}_{-0.15}$	
&050525A$^{a,b}$	
&0.606	
&0.0551$^{+0.0069}_{-0.0062}$	
\\
100302A$^b$	
&4.813	
&0.19$^{+0.20}_{-0.13}$	
&071025$^b$	
&5.2	
&0.3$^{+0.14}_{-0.13}$	
&050505$^b$	
&4.27	
&0.216$^{+0.09}_{-0.156}$	
\\
091208B$^a$	
&1.0633	
&0.0952$^{+0.0098}_{-0.0127}$	
&071010B$^b$	
&0.947	
&0.34$^{+0.12}_{-0.22}$	
&050408$^b$	
&1.2356	
&0.29$^{+0.14}_{-0.16}$	
\\
091127$^b$	
&0.49	
&0.151$^{+0.193}_{-0.071}$	
&071003$^b$	
&1.6044	
&0.3$^{+0.14}_{-0.18}$	
&050315$^{a,b}$	
&1.95	
&0.343$^{+0.038}_{-0.035}$	
\\
091029$^b$	
&2.752	
&0.21$^{+0.19}_{-0.13}$	
&070810A$^b$	
&2.17	
&0.104$^{+0.196}_{-0.049}$	
&050219A$^b$	
&0.2115	
&0.28$^{+0.15}_{-0.16}$	
\\ \cline{7-9}
091024$^b$	
&1.092	
&0.31$^{+0.13}_{-0.15}$	
&070721B$^a$	
&3.6298	
&0.084$^{+0.034}_{-0.029}$	
\\ 
091020$^b$	
&1.71	
&0.28$^{+0.15}_{-0.13}$	
&070714B$^b$	
&0.923	
&0.33$^{+0.11}_{-0.11}$	
\\
091018$^b$	
&0.971	
&0.30$^{+0.14}_{-0.16}$	
&070611$^b$	
&2.0394	
&0.27$^{+0.15}_{-0.11}$	
\\
091003$^b$	
&0.8969	
&0.31$^{+0.13}_{-0.17}$	
&070529$^b$	
&2.4996	
&0.25$^{+0.16}_{-0.12}$	
\\
090902B$^b$ 
&1.822	
&0.31$^{+0.13}_{-0.16}$	
&070521$^b$	
&1.7	
&0.154$^{+0.161}_{-0.079}$	
\\
090812$^b$	
&2.452	
&0.29$^{+0.14}_{-0.15}$	
&070506$^b$	
&2.309	
&0.25$^{+0.17}_{-0.16}$	
\\
090809$^b$	
&2.737	
&0.29$^{+0.13}_{-0.12}$	
&070419A$^b$	
&0.97	
&0.18$^{+0.19}_{-0.11}$	
\\
090726$^b$	
&2.71	
&0.183$^{+0.195}_{-0.094}$	
&070318$^b$	
&0.84	
&0.3$^{+0.11}_{-0.12}$	
\\
090709A$^a$	
&1.8	
&0.287$^{+0.126}_{-0.091}$	
&070306$^a$	
&1.49594	
&0.291$^{+0.03}_{-0.033}$	
\\
090618$^{a,b}$	
&0.54	
&0.059$^{+0.0025}_{-0.0026}$	
&070110$^b$	
&2.3521	
&0.33$^{+0.12}_{-0.15}$	
\\
090519$^b$	
&3.85	
&0.27$^{+0.15}_{-0.16}$	
&070103$^b$	
&2.6208	
&0.29$^{+0.15}_{-0.15}$	
\\
090516A$^a$	
&4.109	
&0.0656$^{+0.0035}_{-0.0045}$	
&061222A$^a$	
&2.088	
&0.0684$^{+0.0142}_{-0.0064}$	
\\
090424$^a$	
&0.544	
&0.218$^{+0.018}_{-0.014}$	
&061126$^{a,b}$	
&1.159	
&0.283$^{+0.126}_{-0.077}$	
\\
090417B$^b$	
&0.345	
&0.183$^{+0.232}_{-0.074}$		
&061121$^b$	
&1.3145	
&0.094$^{+0.118}_{-0.03}$	
\\
090407$^a$	
&1.4485	
&0.3$^{+0.042}_{-0.033}$		
&061021$^b$	
&0.3463	
&0.133$^{+0.06}_{-0.021}$	
\\
\enddata

\tablecomments{$^a$ These are a subsample between 2005 and 2012 (\Swift-Ryan-b) from \citet{rya15}. \\
$^b$These are samples between 2005 and 2012 (\Swift-Ryan 2012) from \citet{rya15}.}
\end{deluxetable}
%
\begin{deluxetable}{|lc|lc|lc||lcc|}
\tabletypesize{\small} %
\tablecaption{{\bf  The \Swift-Perley and the \Swift-Zhang samples}
\label{tbl-4}} %
\tablewidth{-0.pt} %
\tablehead{ %
\colhead{$\rm GRB^{a}$} %
&\colhead{$\rm z$}%
&\colhead{$\rm GRB^{a}$} %
&\colhead{$\rm z$}%
&\colhead{$\rm GRB^{a}$} %
&\colhead{$\rm z$}%
&\colhead{$\rm GRB^{b}$} %
&\colhead{$\rm z$}%
&\colhead{$\rm \theta_{jet}(deg)$}%
} \startdata
050128	
&5.5	
&070223
&1.6295
&090417B
&0.345
&060729
&0.54
&6.15$^{+5.57}_{-1.90}$
\\
050315
&1.95
&070306
&1.4959
&090418A
&1.608
&061121
&1.314
&4.18$^{+0.55}_{-0.19}$
\\
050318
&1.4436
&070318
&0.840
&090424
&0.544
&071020
&2.142
&20.17$^{+0.31}_{-8.26}$
\\
050319
&3.2425
&070328
&2.0627
&090516A
&4.109
&080207
&2.0858
&9.28$^{+7.31}_{-6.50}$
\\
050401
&2.8983
&070419B
&1.9588
&090519
&3.85
&080319B
&0.937
&5.47$^{+21.52}_{-0.08}$
\\
050525A
&0.606
&070508
&0.82
&090530
&1.266
&081007
&0.5295
&28.54$^{+0.12}_{-1.04}$
\\
050726
&3.5
&070521
&2.0865
&090618
&0.54
&090102
&1.547
&11.18$^{+9.20}_{-3.10}$
\\
050730
&3.9693
&070621
&5.5
&090709A
&1.8
&090113
&1.7493
&26.03$^{+2.63}_{-6.63}$
\\
050802
&1.7102
&070721B
&3.6298
&090715B
&3.00
&090417B
&0.345
&28.04$^{+0.62}_{-4.84}$
\\
050803
&4.3
&070808
&1.35
&090812
&2.452
&090423
&8.2
&8.42$^{+12.16}_{-2.73}$
\\
050814
&5.3
&071020
&2.1462
&090814A
&0.696
&091020
&1.71
&8.96$^{+15.58}_{-2.37}$
\\
050820A
&2.6147
&071021
&2.4520
&090926B
&1.24
&091127
&0.490
&26.64$^{+2.02}_{-5.92}$
\\
050822
&1.434
&071025
&4.8
&091018
&0.971
&100615A
&1.398
&22.20$^{+3.18}_{-16.70}$
\\
050904
&6.295
&071112C
&0.8227
&091029
&2.752
&100816A
&0.8049
&28.49$^{+0.17}_{-1.94}$
\\
050922B
&4.9
&080205
&2.72
&091109A
&3.076
&110422A
&1.77
&3.21$^{+0.80}_{-0.61}$
\\
050922C
&2.1995
&080207
&2.0858
&091127
&0.490
&110503A
&1.61
&20.61$^{+1.67}_{-2.10}$
\\
051001
&2.4296
&080210
&2.6419
&091208B
&1.0633
&110731A
&2.83
&27.35$^{+0.10}_{-6.03}$
\\
051006
&1.059
&080310
&2.4274
&100305A
&--
&111008A
&4.9898
&7.32$^{+7.22}_{-1.24}$
\\
060115
&3.5328
&080319A
&2.0265
&100615A
&1.398
&121027A
&1.773
&19.03$^{+6.20}_{-0.25}$
\\ 
060202
&0.785
&080319B
&0.9382
&100621A
&0.542
&130420A
&1.297
&27.13$^{+1.32}_{-0.59}$
\\ \cline{7-9}
060204B
&2.3393
&080325
&1.78
&100728B
&2.106
\\
060210
&3.9122
&080411
&1.0301
&100802A
&3.1
\\
060218
&0.0331
&080413A
&2.4330
&100814A
&1.44
\\
060306
&1.559
&080413B
&1.1014
&110205A
&2.22
\\
060502A
&1.5026
&080430
&0.767
&110709B
&2.09
\\
060510B
&4.9
&080603B
&2.6892
&120119A
&1.728
\\
060522
&5.11
&080605
&1.6403
&120308A
&3.7
\\ \cline{5-6}
060526
&3.2213
&080607
&3.0368
\\
060607A
&3.0749
&080710
&0.8454
\\
060707
&3.4240
&080721
&2.5914
\\
060714
&2.7108
&080804
&2.2045
\\
060719
&1.5320
&080805
&1.5042
\\
060729
&0.5428
&080810
&3.3604
\\
060814
&1.9229
&080916A
&0.6887
\\
060908
&1.8836
&080928
&1.6919
\\
060912A
&0.937
&081008
&1.967
\\
060927
&5.467
&081029
&3.8479
\\
061007
&1.2622
&081109A
&0.9787
\\
061021
&0.3463
&081118
&2.58
\\
061110A
&0.7578
&081121
&2.512
\\
061110B
&3.4344
&081128
&3.4
\\
061121
&1.3145
&081210
&2.0631
\\
061202
&2.253
&081221
&2.26
\\
061222A
&2.088
&081222
&2.77
\\
070110
&2.3521
&090313
&3.375
\\
070129
&2.3384
&090404
&3.0
\\
\enddata

\tablecomments{$^a$These are samples (\Swift-Perley) from the \citet{per16}. \\
$^b$ These are samples (\Swift-Zhang) from \citet{zha15}.}
\end{deluxetable}

To determine the excess of LGRBs at low redshift, in this research we intend to examine the following questions: (i) How does the observed Hubble constant parameter affect the outcomes of the calculated redshift and jet opening angle distributions? In our model, we adopt a $\Lambda$CDM cosmology with the Hubble parameter $H_0 = 72$ km s$^{-1}$ Mpc$^{-1}$, $\Omega_{m} = 0.27$, and $\Omega_{\Lambda} = 0.73$ \citep[e.g.][]{spe03}.  Recently, there has been a disagreement over the value of the Hubble constant; \citet{che17} and \citet{rie16} have suggested that $H_0 = 68$ and $73$ km s$^{-1}$ Mpc$^{-1}$, respectively, and we intend to explore how the Hubble constant $H_0$ affects the calculated distributions. (ii) How does the beaming-corrected gamma-ray energy released relate to redshift? The observed estimated energy release is between $10^{48}$ and $10^{52}$ erg \citep[e.g.][]{fb05}, and the mean beaming corrected gamma-ray energy released for all LGRBs is $\Estarg \approx 2 \times 10^{51}$ erg (see LD07). However, others have also suggested that the average value for \Swift~could be an order of magnitude smaller than for pre\Swift~\citep[e.g.][]{gol16}, which could be due to \Swift's detection of lower energy and higher redshift events. Hence, for self-consistency, these results  need to be considered in the GRB model. (iii) How do the low- and high-energy indices ($\alpha, \beta$) of the Band function \citep[][]{ban93} relate to redshift? Currently, the average accepted values are $\alpha = -1$ and $\beta = -2.5$ for all LGRBs. However, on average, do we see different trends at lower redshift vs higher redshift? The results from these studies will allow us to set the parameter values in our model more appropriately, which will allow us to search for the GRB formation rate in accordance to our model to address the origin of the excess of the LGRBs at low redshift. (iv) Finally, if everything fails to resolve the excess issue, then we need to ask if any of our results point to a possible bias in the \Swift-Ryan LGRB sample? 

Recently, \citet{per16} have deduced a large unbiased \Swift~LGRB sample (hereafter \Swift-Perley) by down selecting the \Swift~sample to remove LGRBs that occurred under circumstances that were not optimal for ground-based follow-up and isolating a subset for which the afterglow redshift completeness is close to the expected maximum achievable value of 80 per cent. This sample is called the \Swift~Gamma-Ray Burst Host Galaxy Legacy Survey (``SHOALS''), a multi-observatory high-redshift galaxy survey targeting the largest unbiased sample of LGRB. This approach was exploited to study unbiased afterglow demographics and redshift distributions by others, including the multiyear Very Large Telescope based ``TOUGH'' project (The Optically Unbiased GRB Host Survey)  \citep[e.g.][and references therein]{per16}. The TOUGH project used a series of observability cuts with bursts that lack the bright afterglows, and isolate a sample of 69 objects that achieves a redshift completeness of 90 per cent, whereas \citet{per16} isolate a sample of 110 objects with 92 percent redshift completeness. The \Swift-Parely sample is better than the \Swift-Ryan-2012 sample because it addresses many possible biases that are affecting the redshift determination besides the afterglow criterial, and they also use GRBs that have bright afterglow observations, while the \Swift-Ryan-2012 sample does not have this tight restriction. We will use \Swift-Perley sample to address possible bias in the \Swift-Ryan 2012 sample, and we refer the readers to \citet{per16} for more details of their sample deduction analysis. The \Swift-Perley LGRB sample is presented in Table~\ref{tbl-4}.

In this paper, we continue to use the complete pre\Swift~\citep[from][]{fb05} and \Swift~samples~\citep[][\Swift-Zhang, \Swift-Ryan,  \Swift-Ryan-b samples]{rya15,zha15} as we have discussed in section~2.3 of LM17 for the purpose of our investigation. As discussed in the LM17 paper, the jet opening angles from the Swift sample are still lacking because of the missing achromatic jet break time \citep[e.g.][]{kz15,zha15}. However, \cite{zha15} were able to estimate the jet opening angles for 27 GRBs from the Swift sample by combining late-time Chandra data with well-sampled Swift/XRT light-curve observations and by fitting the resultant light curves to the numerical simulation assuming an off-axis angle. \citet{rya15}, using an approach similar to that of  \citet{zha15}, have also estimated the jet opening angles of more than 100 GRBs between 2005 and 2012 (\Swift-Ryan 2012 sample). However, only 32 sources (\Swift-Ryan-b) with “well-fitted” jet opening angles were constrained. Interestingly, the \Swift-Ryan-b sample gives a redshift distribution similar to that of the \Swift-Zhang sample, but their mean jet opening angle distribution is much smaller (e.g. much greater than 0.1 rad), as shown in Figure 3(c) of LT17. This result could be related to the fact that the \Swift-Zhang sample is a subset of the \Swift-Ryan 2012 sample, where most of the Chandra-observed GRBs are at the bright end of the whole \Swift~GRB sample \citep[][see Figures 7(c) and 7(d) of LT17]{zha15}. The \Swift-Zhang LGRB jet opening angle sample is presented in Table~\ref{tbl-4}, and the \Swift-Ryan-b sample is as indicated in Table~\ref{tbl-3} along with the complete \Swift-Ryan 2012 sample. In this work, we continue to test how well our model opening angle distribution is correlated with the \Swift-Zhang and \Swift-Ryan-b opening angle samples. In Section~2, we briefly restate the LD07 cosmological GRB model, the redshift and the jet opening angle distributions equations, and our fitting methodology. In Section~3, we discuss the results of our analyses and the possibility of a bias in the \Swift-Ryan 2012 sample, and conclude with a discussion of the derived GRB source rate density and the nature of the low- and high-redshift GRBs in Section~4. We adopt a $\Lambda$CDM cosmology in this paper, where $\Omega_{m} = 0.27$ and $\Omega_{\Lambda} = 0.73$ \citep[e.g.][]{spe03}, while exploring different values of the Hubble constant $H_0$. 

\newpage
\section{A BRIEF STATEMENT OF OUR COSMOLOGICAL GRBs MODEL}
In this section, we give a very brief overview of the LD07 model, the bursting rate of GRB sources, and our fitting methodology, and refer the readers to their paper for more details. To calculate the redshift, size, and the jet opening angle distributions, we use Equations (16), (18), and (20) from LD07, which describe the directional GRB rate per unit redshift ($d\dot{N}(> \fluxthres)/d\Omega dz$) (or the redshift distribution) with energy flux $>\fluxthres$, the size distribution of GRBs ($d\dot{N}(> \fluxthres)/d\Omega$) in terms of their $\nu F_\nu$ flux $f_\e$, and the observed directional event reduction rate (due to the finite jet opening angle) for bursting sources with $\nu F_\nu$ spectral flux greater than $\fluxthres$ at observed photon energy $\e$ or simply the jet opening angle distribution ($d\dot{N}(> \fluxthres)/d\Omega d\mu_{\rm j}$), respectively. These three equations require the jet opening angle distribution $g(\mu_{\rm j})$, the comoving GRB rate density $\dot{n}_{co}(z)$, and the instrument's detector sensitivity $\fluxthres$. In this paper, we continue to use $\fluxthres \sim 10^{-8} $ and $\sim 10^{-7}$ erg cm$^{-2}$ s$^{-1}$ (see LD07) as the effective \Swift~and pre\Swift~detective flux thresholds, respectively. Since the form for the jet opening angle $g(\mu_{\rm j})$ is unknown, we also consider the function
$g(\mu_{\rm j}) = g_0 \ (1-\mu_{\rm j})^s \ H(\mu_{\rm j};\mu_{\rm j,min},\mu_{\rm j,max})$,
where $s$ is the jet opening angle power-law index; for a two-sided jet, $\mu_{\rm j,min} \geq 0$, and 
$g_0 = (1+s)/((1-\mu_{\rm j,min})^{1+s} - (1-\mu_{\rm j,max})^{1+s})$ 
is the distribution normalization (LD07), and $H(\mu; \mu_{\rm j}, 1)$ is the heaviside function such that $H(\mu; \mu_{\rm j}, 1) = 1$ when $\mu_{\rm j} \le \mu \le1$ (or when the angle $\theta$ of the observer with respect to the jet axis is within the opening angle of the jet), and $H(\mu; \mu_{\rm j}, 1) = 0$ otherwise. According to the observations from the pre\Swift~and \Swift~instruments, the minimum and maximum jet opening angles are about $\theta_{\rm j,min} \sim 0.042$ and $\theta_{\rm j,max} \sim 0.7$ rad, where $\mu_{j} = cos(\theta_j)$~\citep[e.g.][]{fb05,rya15,zha15}. This functional form $g(\mu_{\rm j})$ describes GRBs with small opening angles, $\theta_{\rm j} \ll 1$, so that such GRBs are potentially detectable from larger distances, for the same energy budget. By contrast, GRB jets with large opening angles are more frequent, but only detectable from comparatively small distances (e.g. LD07). The possibility of beaming angle being narrower at higher redshift has also been suggested by others \citep[e.g.][]{lfr02,lu12,las14,las18a,las18b,laj19,lha19}.
These requirements require that the maximum redshift $z_{max}$ must satisfy the condition
$d^2_L(\zmax) \; \leq \; \Estarg/(4 {\rm \pi} (1 - \mu_{\rm j,min}) \; \Delta t_* \; \fluxthres \; \lambda_b)$, where $d_L(z) = \frac{c}{H_0}(1+z) \ \int^z_0 \frac{dz^\prime}{\sqrt{\Omega_m (1+z^\prime)^3 + \Omega_\Lambda}}$ is the luminosity distance, $\Delta t_*$ is the duration of the GRB in the stationary frame and we set it equal to $10 s$ as the average observed value based on BATSE GRBs at $z \sim 1$, and $\lambda_b$ is the bolometric correction to the peak measured $\nu F_\nu$ flux as described by Equations (1) and (2) in LD07 with $\lambda_b = (a^{-1} - b^{-1})$, where $a$ and $b$ are the broken power-law indices of the GRB spectrum. We refer the readers to section 2 of LD07 for more details of our model.
Finally, to close the system, the comoving GRB rate density is defined as
$\dot{n}_{co}(z) = \dot{n}_{co} \Sigma_{_{\rm SFR}}(z)$,
where $\dot{n}_{co}$ is the normalization constant and $\Sigma_{_{\rm SFR}}$ is the GRB formation rate as a function of redshift of the form
\begin{equation}
    \Sigma_{_{\rm SFR}}(z)=\left\{
                \begin{array}{ll}
                  a_0 (1+z)^{\eta_{_1}} \hskip+0.25in , 0 \leq z \leq z_1 \\
                  b_0 (1+z)^{\eta_{_2}} \hskip+0.25in , z_1 < z \leq z_2  \;, \\
                  c_0 (1+z)^{\eta_{_3}} \hskip+0.25in , z > z_2 
                \end{array}
              \right.  
\label{eq1}
\end{equation}
where $a_0 = 1$, $b_0 = a_0 (1+z_1)^{\eta_{_1}} / (1+z_1)^{\eta_{_2}}$, and $c_0 = b_0 (1+z_2)^{\eta_{_2}} / (1+z_2)^{\eta_{_3}}$, and $\eta_{_1}, \eta_{_2}$, $\eta_{_3}$ are constants, and the redshift values $z_1$ and $z_2$ are equal to $0.5$ and $4.5$, respectively.  We select these breaking redshift values because they provide the best fit to the observed distributions as discussed in LM17 for SFR9, and we continue to use them here in this paper.  

The constant values $\eta_{_1}, \eta_{_2}$, $\eta_{_3}$, and the power-law index $s$ will be fine tuned until the calculated model distributions give the best fit to the observed pre\Swift~and \Swift~redshift and jet opening angle distributions for a fix set of the physical parameters values ($H_0, a, b, \theta_{\rm j,min}, \theta_{\rm j,max}, \Estarg,$ and $\fluxthres$). If the model fails to fit the data, then we have to readjust the physical parameters values and refit the model to the data. It is important for the readers to realize that for us to accept any of our parameter values to be the best values for any fits, we require those values to give the best fit to the redshift and jet opening angle distributions for both the \Swift~and pre\Swift~data at the same time. This allows us to quantify how good a fit is and the actual preferred parameters of the model for the given data. We then perform a KS-test as a final check to validate that the observed data sets are drawn from the calculated model distribution by examining the $D$-statistic and the probability $p$-value. We refer the readers to the LD07 paper in section~3.2 for the details of our fitting analysis. Once we have obtained the best functions for the redshift and jet opening angle model distributions for both the pre\Swift~and \Swift~data, we will then calculate the count rates and the size distribution (Equation (18) from LD07) from our model and test them against the expected results for the \Swift~and BATSE instruments. 
\\
\begfig[t] \hskip-0.25in \epsscale{1.15} \plottwo{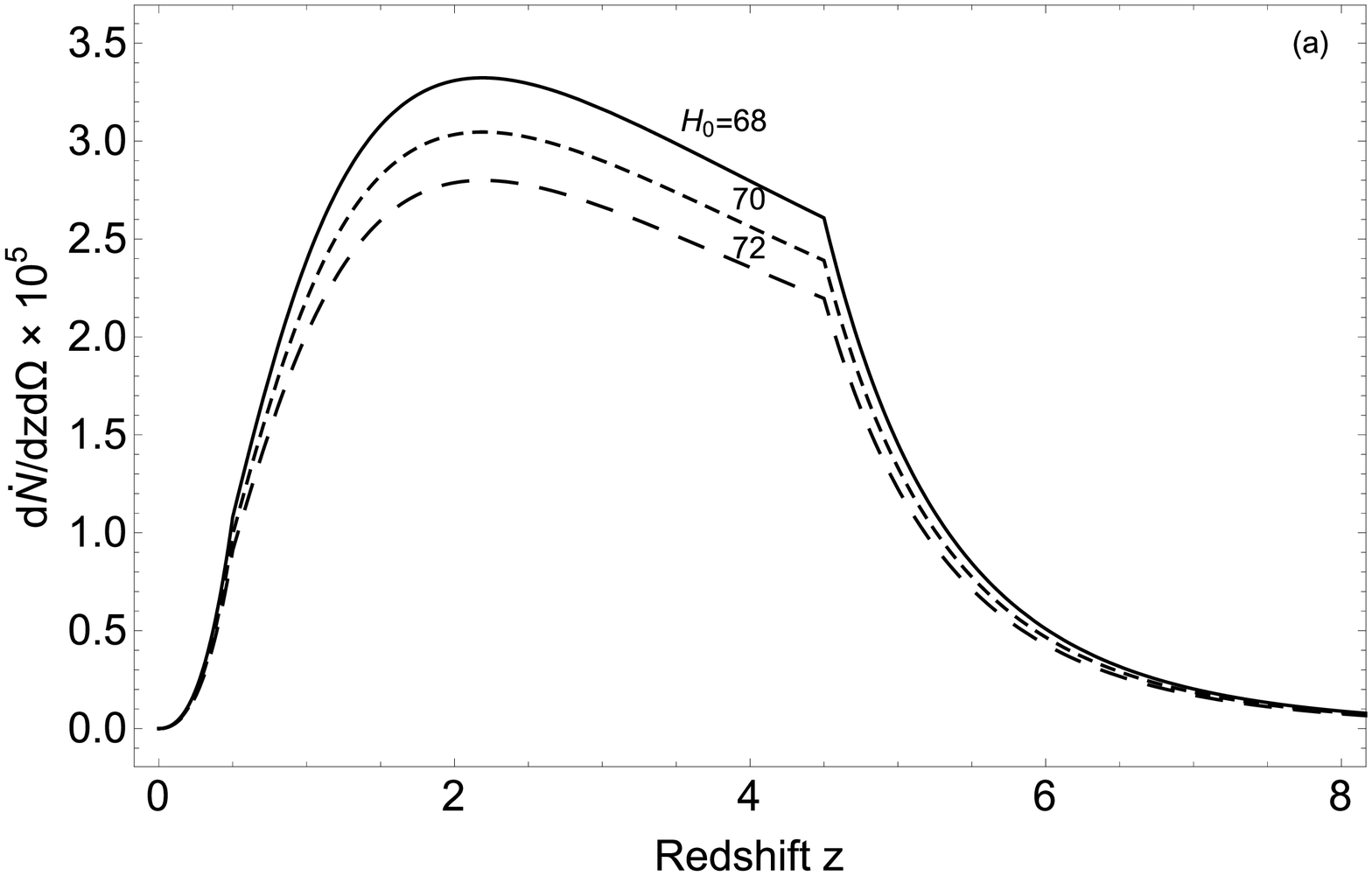}{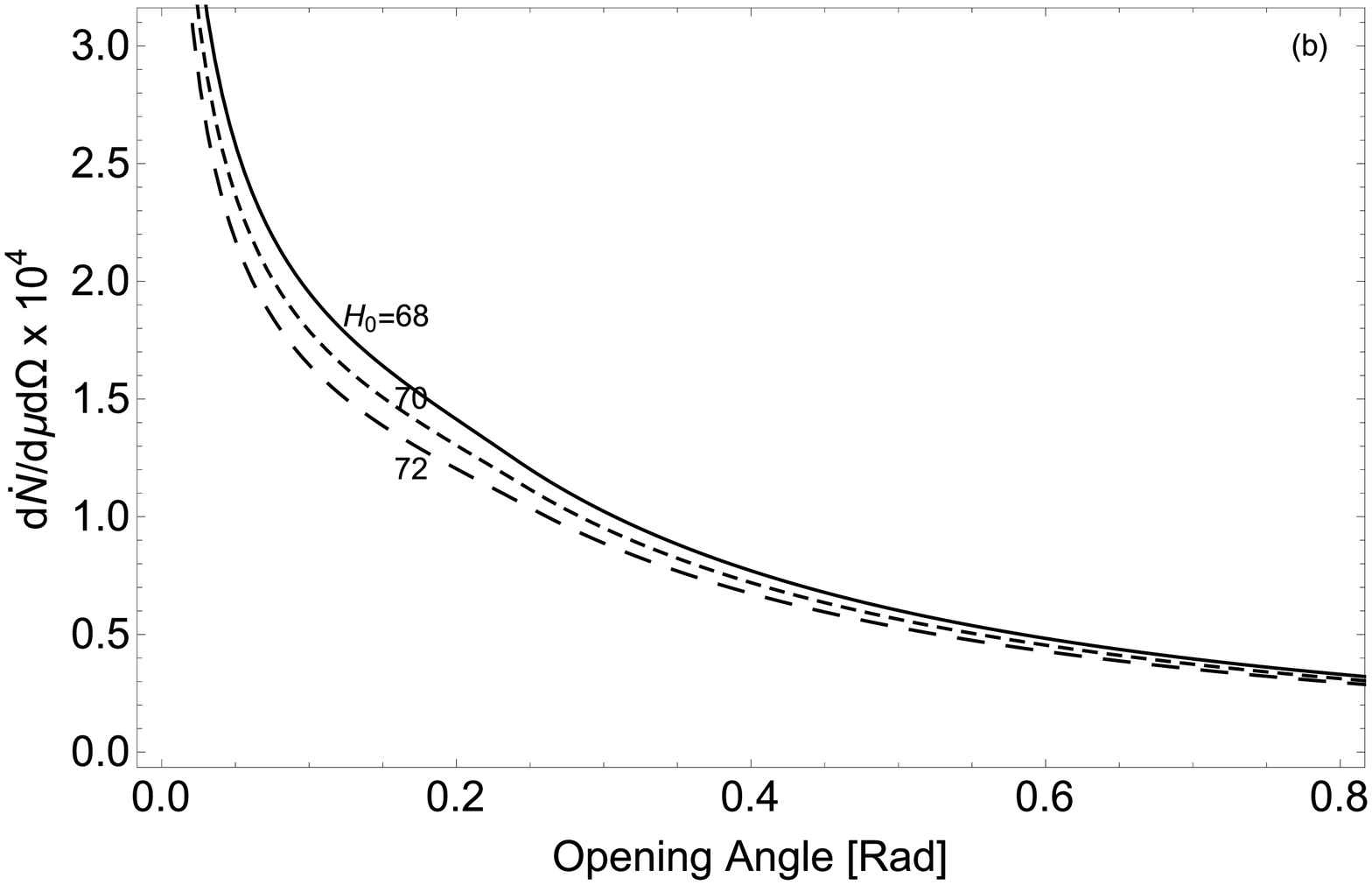}
\caption{\footnotesize ({\it a}) Directional event rate per unit redshift [one event per $(10^{28} \ \rm cm)^3$ per day per $sr$ per $z$] and ({\it b}) directional event rate per unit jet opening angle [one event per $(10^{28} \ \rm cm)^3$ per day per $sr$ per $\mu_{\rm j}$] using $\theta_{\rm j,min} = 0.05 \ \rm rad$ and $\theta_{\rm j,max} = 0.7 \ \rm rad $,  $s = -1.2$, $\Estarg = 2 \times 10^{51} \ \rm erg$, $\fluxthres = 10^{-8} \ \rm erg \ cm^{-2} \ s^{-1}$,  $\lambda_b = 3$, and SFR9. The solid, short, and long dash curves represent the Hubble constant values of $H_0 = 68, 70$, and $72$ km s$^{-1}$ Mpc$^{-1}$, respectively.}
\label{fig2} 
\finfig
\begfig[t] \hskip-0.25in \epsscale{1.15} \plottwo{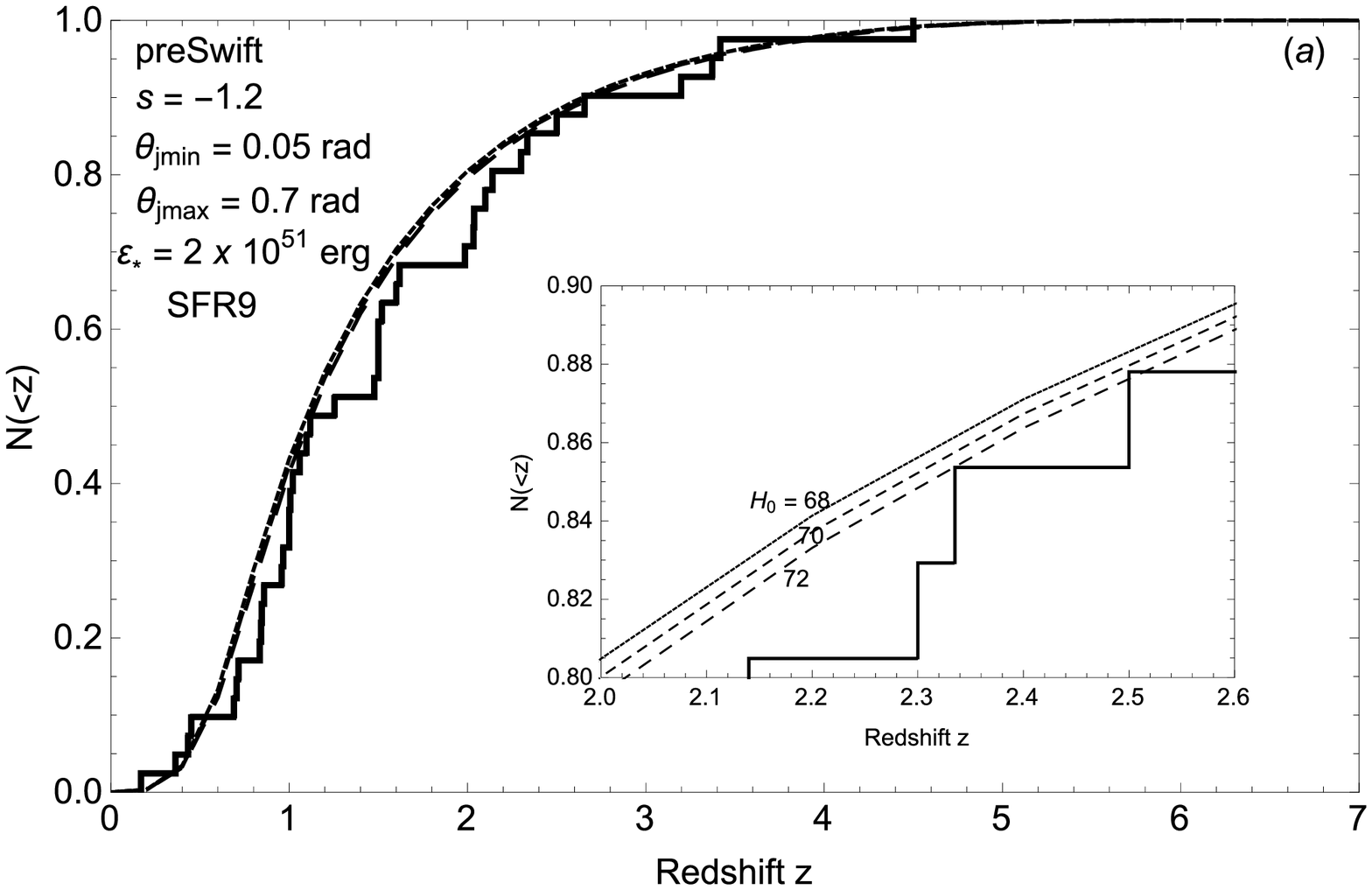}{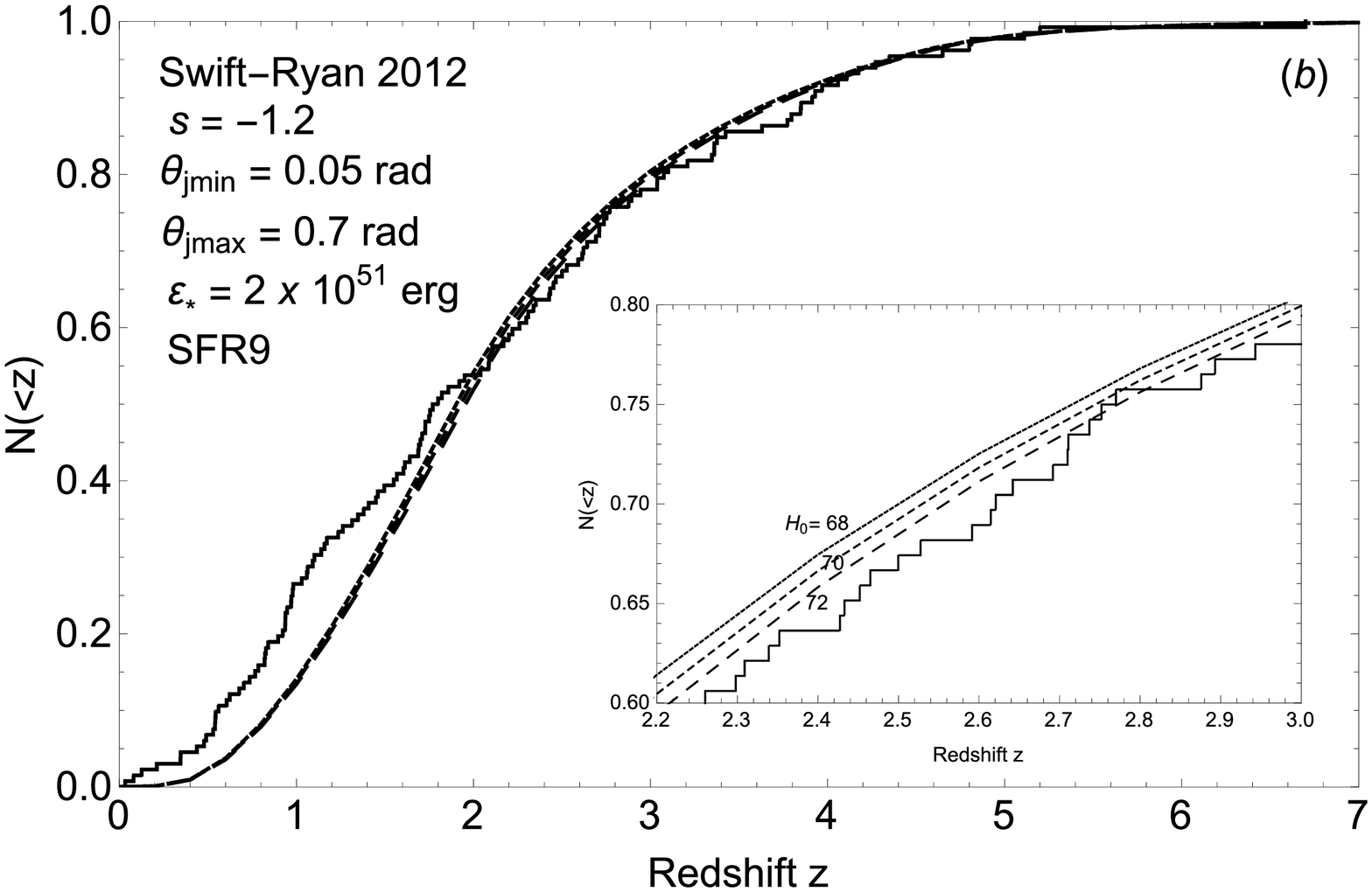}
\caption{\footnotesize ({\it a}) pre\Swift~and ({\it b}) \Swift~redshift distribution using $\theta_{\rm j,min} = 0.05 \ \rm rad$ and $\theta_{\rm j,max} = 0.7 \ \rm rad $,  $s = -1.2$, $\Estarg = 2 \times 10^{51} \ \rm erg$, $\fluxthres = 10^{-7} \ \rm erg \ cm^{-2} \ s^{-1}$ (pre\Swift), $\fluxthres = 10^{-8} \ \rm erg \ cm^{-2} \ s^{-1}$ (\Swift),  $\lambda_b = 3$, and SFR9. The insets in Figures 3(a) and 3(b) are the same plots as Figures 3(a) and 3(b), respectively, but at a small range of redshift for clarity. The solid curves are the complete pre\Swift-Friedman \& Bloom and \Swift-Ryan 2012 redshift distributions. The dotted, short, and long dash curves are the calculated distribution using the Hubble constant values of $H_0 = 68, 70$, and $72$ km s$^{-1}$ Mpc$^{-1}$, respectively.}
\label{fig3} 
\finfig
\section{RESULTS AND DISCUSSION}
Similar to the work done by LD07 and LM17, there are seven physical adjustable parameters in this model: the $\nu F_\nu$ spectral power-law indices $a$ and $b$, the power-law index $s$ of the jet opening-angle distribution, the range of the jet opening angles $\theta_{\rm j,min}$ and $\theta_{\rm j,max}$, the average absolute emitted gamma-ray energy $\Estarg$, and the detector threshold $\fluxthres$ excluding the Hubble constant $H_0$ and the constant values of $\eta_{_1}, \eta_{_2}$, and $\eta_{_3}$ of the GRB rate density. The flux thresholds $\fluxthres$ are set equal to $10^{-8}$  and $10^{-7}$ erg cm$^{-2}$ s$^{-1}$ for \Swift~and pre\Swift, respectively, and the observed minimum and maximum jet opening angles are constrained by $\theta_{\rm j,min} = 0.042$ and $\theta_{\rm j,max} = 0.7$ rad, respectively.  As mentioned, if we consider a broken power-law $\nu F_\nu$ GRB SED with $\alpha=-1$ and $\beta = -2.5$, which give $a =1$ and  $b = -0.5$ as the general accepted values, then this gives the bolometric correction factor $\lambda_b = 3$.  However, in this paper we will investigate for the best values of $\alpha$ and $\beta$ as a function of redshift based on the observed data. The remaining parameters $s$, $\Estarg$, and the GRB rate functional form are constrained by obtaining the best model distributions that represent the observed redshift and jet opening angle distributions for both \Swift~and pre\Swift~instruments.

We go through point-by-point addressing the concerns that we have mentioned in Section~1. From our model, Figures~\ref{fig2}(a)-\ref{fig2}(b) are the plots of the directional event rate per unit redshift per steradian and directional event rate per unit jet opening angle per steradian for different values of the Hubble constant $H_0 = 68, 70$, and $72$ km s$^{-1}$ Mpc$^{-1}$. As we increase the value of the Hubble constant, the luminosity distance decreases and, consequently, the directional event rate per unit redshift per steradian and directional event rate per unit jet opening angle per steradian are reduced over all redshift and jet opening angle as depicted in Figures~\ref{fig2}(a) and \ref{fig2}(b) as according to equations (16) and (20) in LD07. These results suggest the overall decrease in the total number of GRBs as $H_0$ increases as indicated in Figures~\ref{fig2}(a)-\ref{fig2}(b); and this is expected since large value of $H_0$ implies a younger Universe. The results in Figures~\ref{fig3}(a) and \ref{fig3}(b) further suggest that the redshift distributions, between our model and the pre\Swift~and \Swift-Ryan 2012 sample, are not sensitive to the value of the Hubble constant. However, it is interesting to point out that the directional event rate per unit redshift and jet opening angle in Figures~\ref{fig2}(a) and \ref{fig2}(b) seem to suggest otherwise. This indicates that it is more advantageous to analyze the observed redshift or the jet opening angle as directional event rate distribution. We plan to explore this result in future paper. For the rest of this paper, we will use $H_0 = 72$ km s$^{-1}$ Mpc$^{-1}$.
\begfig[t] \hskip-0.25in \epsscale{1.15} \plottwo{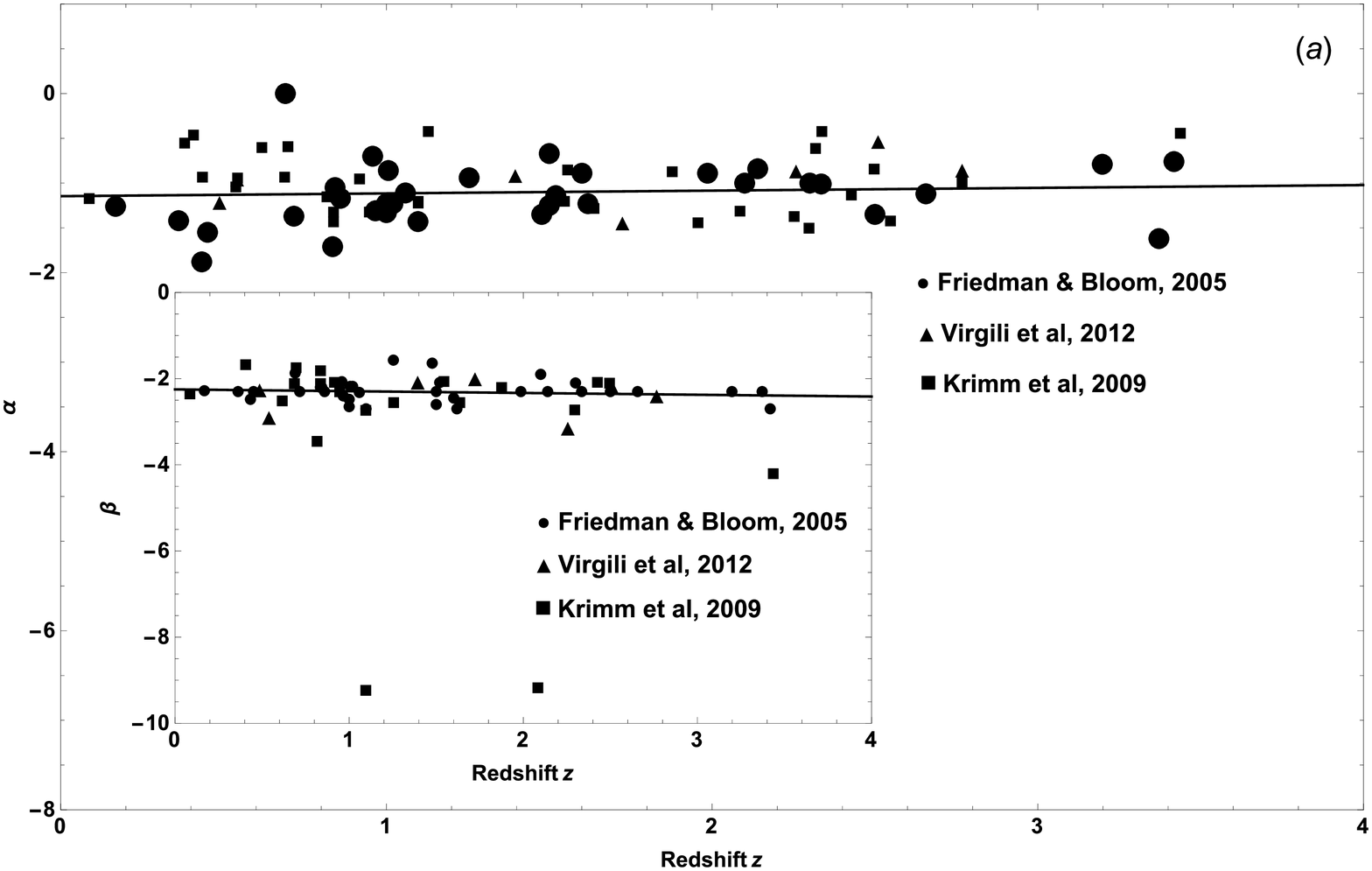}{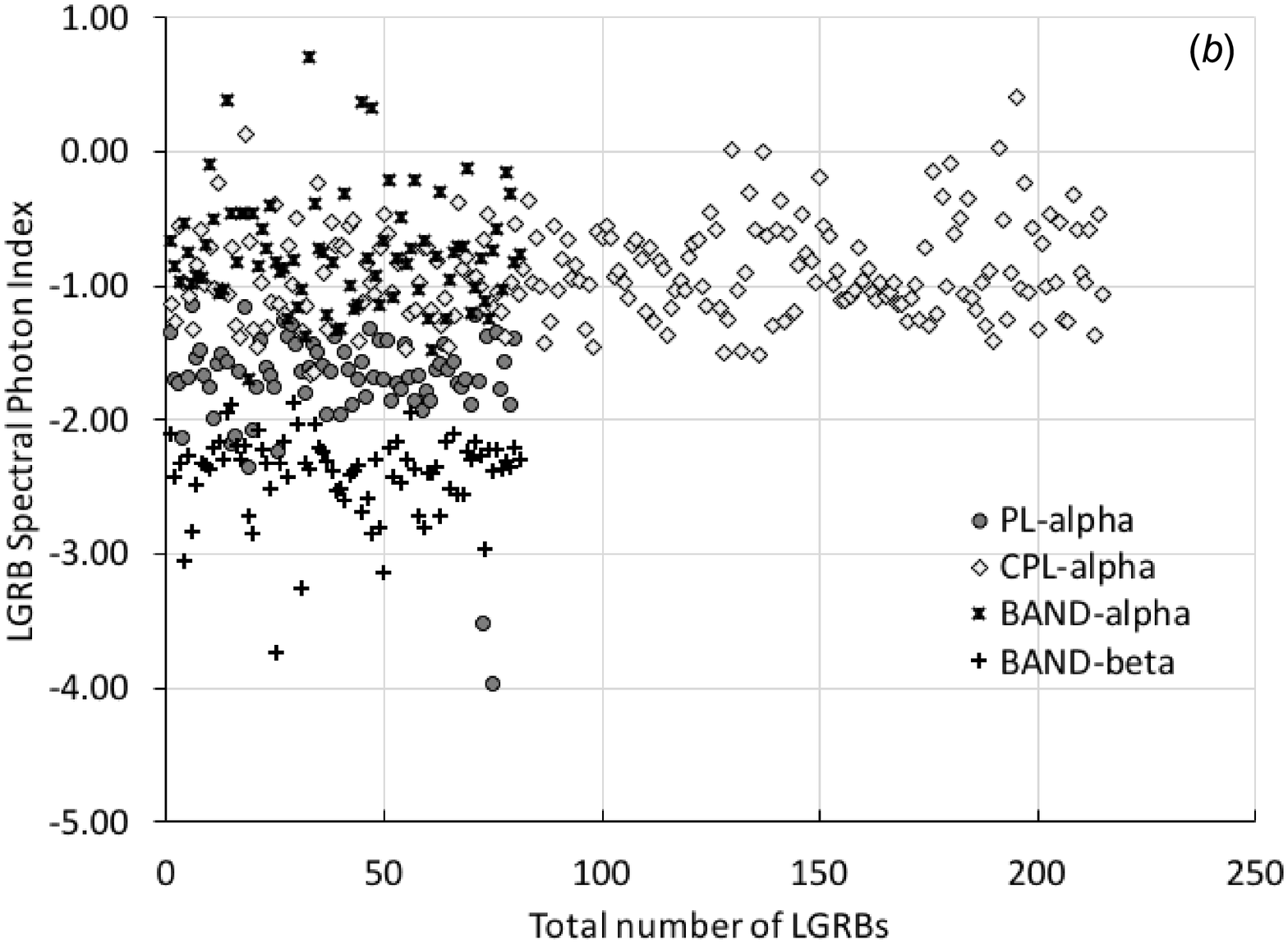}
\caption{\footnotesize ({\it a}) These are the average low-energy ($\alpha$) and the average high-energy ($\beta$) photon indices, inset in Figure 4(a), of the GRB spectra for different GRBs at different redshift that based on the phenomenological model of the Band function as observed by pre\Swift, \Swift, and \Fermi. The data (the solid circles, triangles, and squares) are taken from \citet{fb05}, \citet{kri09}, and \citet{vir12}, respectively. The solid lines from both graphs are the best-fitting lines that show a flat fit through the data. The two data points from \citet{kri09} near $\beta \sim -9$ at redshifts of 1 and 2 are outliers, and the best-fitting line does not take these two values into account. ({\it b}) These are the low-energy ($\alpha$) and high-energy ($\beta$) photon indices of the GRB spectra for different GRBs at different redshift that based on the phenomenological model as observed by \Fermi. The solid circles, rhombuses, x, and plus represent the photon indices of the GRB spectral using the smooth power law, cut-off power law, and the $\alpha$ and $\beta$-Band function, respectively. These data are taken from \citet{nav11}.}
\label{fig4} 
\finfig

The plots in Figures~\ref{fig4}(a)-\ref{fig4}(b) are the average values of the low- and high-energy indices, $\alpha$ and $\beta$, of the Band function of different GRB at different redshift as observed by pre\Swift, \Swift, and \Fermi~\citep[the data are obtained from][]{fb05,kri09,vir12}.  From Figure~\ref{fig4}(a), the $\alpha$ and $\beta$ (inset of Figure~\ref{fig4}(a)) values are independent of redshift with the average values of about $-2.5$ and $-1$, respectively. Using \Fermi~data, \citet{nav11} did a spectral analysis study and their results indicate that either the Band model or the Comptonized model (power law with a high-energy exponential cutoff) best represents the long GRBs with the average values of $\alpha \approx -0.9$ and $\beta \approx -2.3$ as depicted in Figure~\ref{fig4}(b), and all three plots show large scattering about the mean. 

Figures~\ref{fig5}(a)-\ref{fig5}(c) contain the parameter values that best represent both the pre\Swift~and \Swift~redshift and jet opening angle model distributions with the low- and high-energy indices values of $\alpha \approx -0.9$ and $\beta \approx -2.18$, and the GRB formation rate density SFR11 as indicated in Figure~\ref{fig1}(a). Note that SFR11 has a much higher GRB density rate below a redshift of 2 than SFR9, and its model parameter values are indicated in Table~\ref{tbl-2}. Furthermore, the low- and high-energy indices values are also within the mean values as discussed above.  Nevertheless, the excess of LGRBs below a redshift of 2 is still present as shown in Figure~\ref{fig5}(b) (an inset of Figure \ref{fig5}(a)), but a big improvement to the overall curve is achieved in comparison to the result in Figure~\ref{fig1}(b) using SFR9 model. This implies that there is something else affecting the LGRBs redshift distribution, unless, our sample is contaminated, for example, with an intermediate LGRBs duration \citep[e.g.,][]{muk98,ver10}, for which the physical mechanism is unclear (can be off-axis, high baryon load GRBs etc.).  

Table~7 from \citet{ver10} paper contains 24 intermediate type bursts. We cross-check with our sample, which are restricted to $T_{90} > 2$ s, and find four intermediate bursts in our sample (GRBs: 050824, 051016B, 060926, and 081007). Nevertheless, the redshift distribution shows very little variation after removing these four intermediate bursts from our sample. Moreover, it is interesting to note that our results provide a good fit to the \Swift-Ryan-b sample and not to the \Swift-Zhang sample for both the redshift and jet opening angle distributions (see the insets of Figures \ref{fig5}(b) and \ref{fig5}(c), respectively). We do not plot the \Swift-Zhang redshift distribution here because we are only interested in its opening angle distribution. The \Swift-Zhang redshift distribution is shown in Figure~3(b) of LM17. The shaded regions in the jet opening angle distribution plot represent the error bars from the \Swift-Zhang and \Swift-Ryan-b samples. We also notice the \Swift~calculated jet opening angle distribution is within the error margin of the \Swift-Ryan jet opening angle sample (see Figure 7c in LB17). It is important to remind the readers that the \Swift-Ryan-b sample is a subset of the \Swift-Ryan-2012, where only 32 sources with ``well-fitted'' jet opening angles are constrained \citep[see][]{rya15}.

\begfig[t] \hskip-0.25in \epsscale{1.15} \plottwo{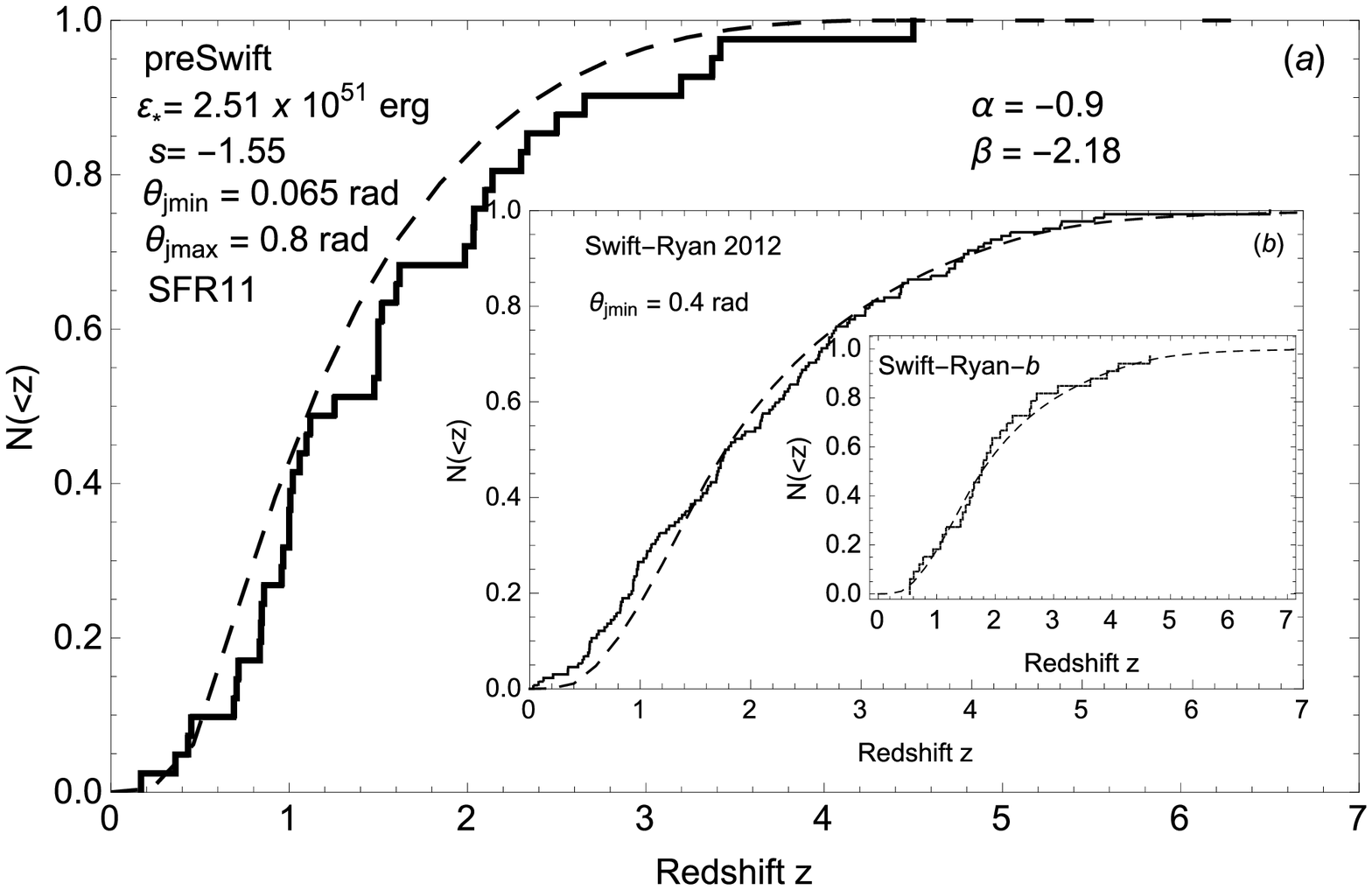}{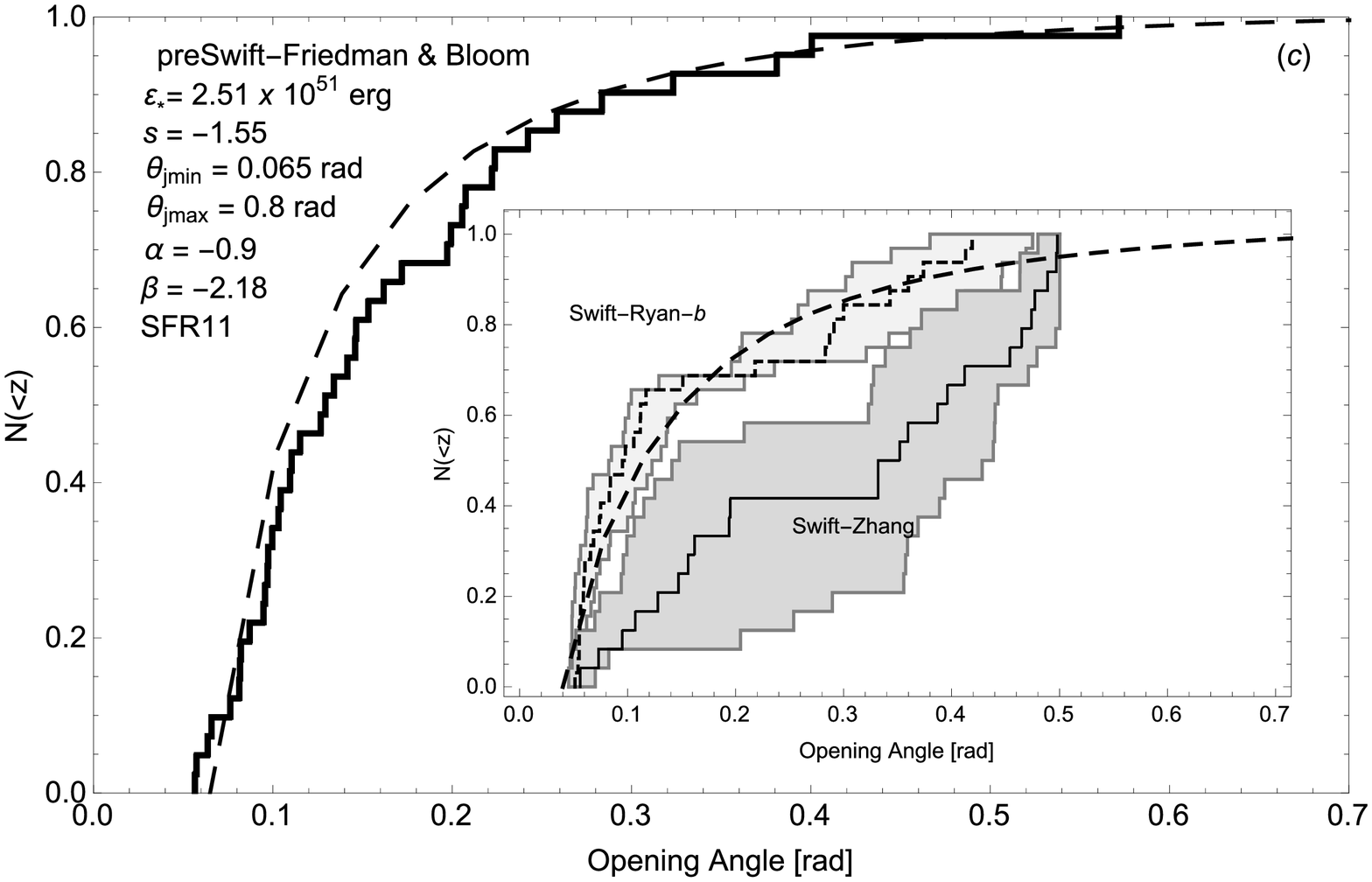}
\caption{\footnotesize ({\it a}) These are the pre\Swift~redshift and ({\it c}) jet opening angle distributions. Figure 5(b) and the insets of Figures 5(b) and 5(c) are similar as plots in Figures 5(a) and 5(c), respectively, but for the \Swift~data. Both the pre\Swift~and \Swift~model distributions share the same parameter values ($\theta_{\rm j,min} = 0.065 \ \rm rad$ and $\theta_{\rm j,max} = 0.8 \ \rm rad $,  $s = -1.55$, $\Estarg = 2.51 \times 10^{51} \ \rm erg$,  $\alpha = -0.9$, $\beta = -2.18$, Hubble constant of $H_0$ = $72$ km s$^{-1}$ Mpc$^{-1}$, and SFR11), except, $\fluxthres = 10^{-7} \ \rm erg \ cm^{-2} \ s^{-1}$ and $\fluxthres = 10^{-8} \ \rm erg \ cm^{-2} \ s^{-1}$ for pre\Swift~and \Swift, respectively. The solid curves in Figures 5(a) and 5(b) are the complete pre\Swift-Friedman \& Bloom and \Swift-Ryan 2012 redshift distributions. The solid curve in the inset of Figure 5(b) is the subsample redshift distribution from the \Swift-Ryan-2012 sample. The solid curves in Figure 5(c) and the solid curves of the \Swift-Zhang sample and the short dashed curve of the \Swift-Ryan-b sample in the inset of Figure 5(c) are the observed opening angle distributions. The shaded regions of the opening angle distribution plot in the inset of Figure 5(c) represent the error bars from the \Swift-Zhang and \Swift-Ryan-b samples. The long dash curves in all figures are the calculated distribution. The parameter values in Figures~5(a) and 5(c) also apply to their insets.}
\label{fig5} 
\finfig

\citet{lie16} and \citet{sak11} showed that most of the GRB spectra in the BAT energy range can be well fitted by the simple power-law model of the photon flux with a spectral index $\alpha_{\rm pl}$. Figure~\ref{fig6}(a)-\ref{fig6}(c) are the plots for the pre\Swift~and~\Swift~redshift and jet opening angle distributions, where $\alpha_{\rm pl} = -1.28$ and SFR10 are the value and model that provide the best pre\Swift~and \Swift~model distributions. According to our model, this would require the bolometric correction to be $\lambda_b = 1/a$, where $a = 2 + \alpha_{\rm pl}$. The inset of Figure~\ref{fig6}(b) is a plot of the spectra index $\alpha_{\rm pl}$ for the simple power law as function of redshift, and the data is obtained from \citet{lie16} paper. The plot indicates that $\alpha_{\rm pl}$ is independent of redshift with an averaged value of $\alpha_{\rm pl} \sim -1.5$, and our best value is within $1 \sigma$ of the observed distribution. Using the  parameter values as indicated in the figure, the results show an improvement to every distributions in comparison to Fig.~\ref{fig5}, except the \Swift~jet opening angle distribution (see the inset of Fig.~\ref{fig6}c); the plot shows that the \Swift~jet opening angle model  distribution (the long dashed curve) is outside the error bars of the observed jet opening angle distribution. Moreover, Fig.~\ref{fig6}(b) still indicate an excess of LGRB below a redshift of 2.

To check which LGRB density rate model, SFR10 or SFR11, is more favourable, we compare our model size distribution with BATSE 4B Catalogue size distribution using SFR10 and SFR11 for the simple power-law and the broken power-law GRB spectral, respectively. BATSE 4B Catalogue contains 1292 bursts total, including short and long duration GRBs. There are a total of 872 LGRBs that are identifiable in the BATSE 4B Catalogue \citep{pac99,ld09} with a BATSE detection rate of $\approx 550$ GRBs per year full-sky brighter than 0.3 photons $\rm cm^{-2} \, s^{-1}$ in the $50-300$ keV band \citep{ban02}. Since there are only 872 LGRBs in the catalogue, hence, BATSE detection rate is $\approx 371$ LGRBs per year full sky. In Figure~\ref{fig7}(a), we plot the differential size distribution of BATSE GRBs using the Fourth BATSE catalogue \citep{pac99} in comparison with our model prediction. The plots are normalized to the current total number of observed LGRBs per year from BATSE, which is 371 bursts per $4 \pi$ sr exceeding a peak flux of 0.3 photons $\rm cm^{-2} \ s^{-1}$  in the $\Delta E=50-300$ keV band for the $\Delta t=1.024$ s trigger time. As can be seen from Figure~\ref{fig7}(a), the broken power-law model gives a good  agreement to the differential burst rate, while the simple power-law model overpredicts the differential burst rate above $1.0 \ \rm ph \ cm^{-2} \ s^{-1}$. It is important to note that the size distribution of the BATSE GRB distribution in Figure~\ref{fig7}(a) is normalized and corrected to 371 LGRBs above a photon number threshold of $0.3 \ \rm ph \ cm^{-2} \ s^{-1}$. Below a photon number threshold of $0.3 \ \rm ph \ cm^{-2} \ s^{-1}$ in the 50 -- 300 keV band, the observed number of GRBs falls rapidly due to the sharp decline in the BATSE trigger efficiency at these photon fluxes \citep[see][]{pac99}, while the size distribution of the \Swift~GRBs will extend to much lower values $\cong 0.0633 \ \rm ph \ cm^{-2} \ s^{-1}$. The result in Figure~\ref{fig7}(a) indicates that SFR11 gives an excellent fit to the BATSE LGRB, while SFR10 overpredicts the differential size distribution above $1.0 \ \rm ph \ cm^{-2} \ s^{-1}$.

Additionally, in Figure~\ref{fig7}(b), we recalculate the pre\Swift~model distributions using $\Estarg = 4.47 \times 10^{51} \ \rm erg$ for the SFR11 model; this value is fine tuned, while holding the other physical parameter values the same as indicated in Figures~\ref{fig5}(a) and \ref{fig5}(c), until the redshift and jet opening angle model distributions provide the best fit to the observed pre\Swift~data. We notice that the curves to the pre\Swift~redshift and jet opening angle distributions are greatly improved, and this result implies that the mean beaming corrected gamma-ray energy released for the pre\Swift~is about a factor of 2 larger than the \Swift~sample. This result mostly agrees with \citet{gol16} analysis, however, instead of being off by an order of magnitude our result indicates that it is off by a factor of 2. The nature of the difference is that we obtain our value based on fitting the redshift and jet opening angle distributions from both the pre\Swift~and \Swift~samples, while \citet{gol16} based on estimating the {\it Fermi}-GBM (Gamma-ray Burst Monitor) jet opening angle samples that relies on many factors including the GRBs with observed and unobserved redshifts as indicated in Eq.(5) of their paper. \citet{gol16} study a sample of 638 long GRBs with only 40 observed redshifts. Their estimates of $\theta_j$ are consistent with the calculation of $\theta_j$ via the observed jet breaks (see their Eq. (3)) within $1\sigma$ confidence level for the observed redshift, while the estimates of $\theta_j$ have large uncertainty of $\theta_j$ when the redshift is unknown. Their $\theta_j$ distribution shows that the majority of the jet peaks at, $< 10^\circ$, and our angle distribution indicates that the majority of the jet occurs around $\sim 7^\circ$ as indicated in the inset of Figure~\ref{fig5}(b) for the \Swift~data, and this is consistent with their result. However, since their GRBs jet opening angle peak around $10^\circ$, according to our work, this will reduce the estimate of the collimation-corrected energy $\Estarg$. We refer the readers to their paper for more details of their analysis. This factor of 2 suggests that on average the LGRBs that were detected at redshift less than 4 have larger gamma-ray energy released than the LGRBs that were detected at redshift greater than 4. If this is the case, then our \Swift~model distribution would require two different values of gamma-ray energy released separated at a redshift of 4. Based on our model,  this will cause the \Swift~model redshift and jet opening angle distributions to shift to a higher redshift and a lower opening angle, respectively. Consequently, this would provide a bad fit to both the observed redshift and opening angle distributions from the \Swift-Ryan and \Swift-Ryan-b samples. Nevertheless, it is important to notice that our model redshift distributions (see Figure~\ref{fig5}(b)) still indicate that there is an excess of LGRB at redshift less than 2 in the \Swift-Ryan 2012 sample using SFR11 model. 

From the above analyses, SFR11 seems to provide the best model for the LGRB density rate with a broken power-law $\nu F_\nu$ GRB spectrum that resulted in either an excess of LGRB below a redshift of 2 as shown in Figure~\ref{fig5}(b) based on the \Swift-Ryan 2012 sample or an absence of the excess as shown in the inset of Figure~\ref{fig5}(b) based on the \Swift-Ryan-b sample. It is also important to remind the readers that the SFR11 model also gives a good fit to the observed \Swift-Ryan-b and pre\Swift~and jet opening angle distributions as indicated in the inset of Figures~\ref{fig5}(c) and~\ref{fig7}(b), respectively. Since the calculated distributions seem to correlate well with the \Swift-Ryan-b redshift and jet opening angle distributions and as well as the pre\Swift~distributions, this could indicate that there is a biased in the \Swift-Ryan sample.   
\begfig[t] \hskip-0.25in \epsscale{1.15} \plottwo{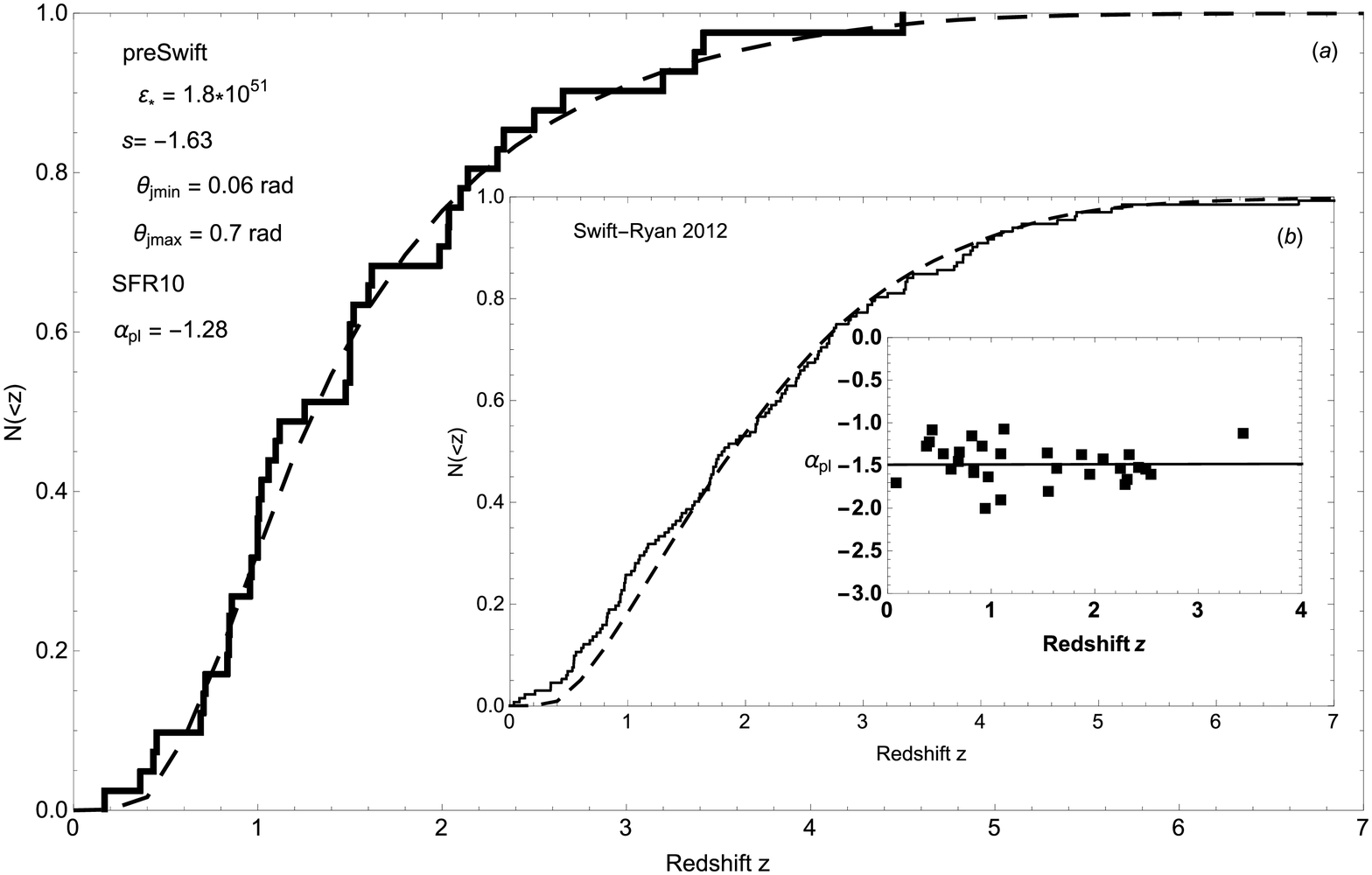}{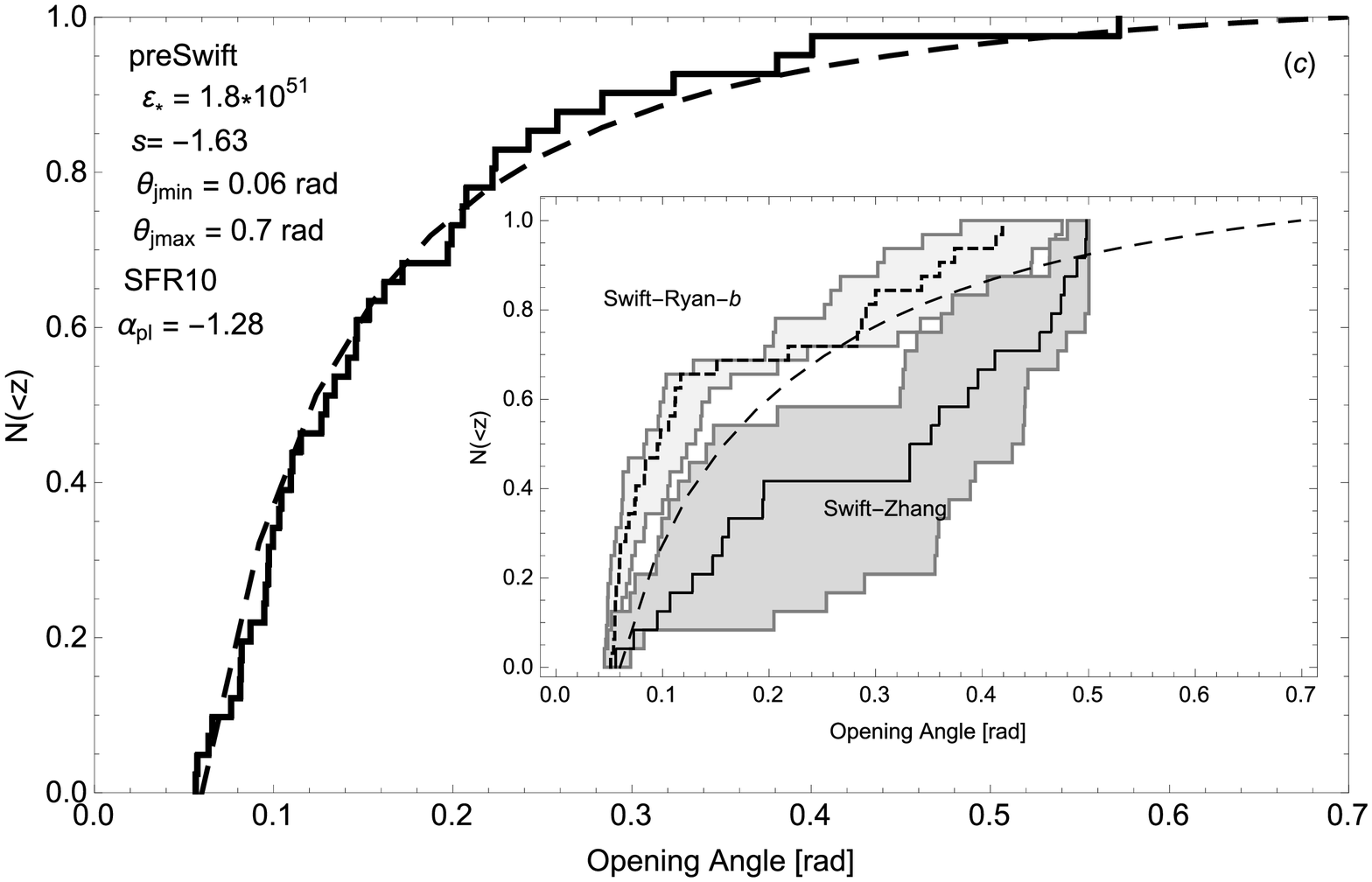}
\caption{\footnotesize ({\it a}) pre\Swift~redshift and ({\it c}) pre\Swift~jet opening angle distributions using $\theta_{\rm j,min} = 0.06 \ \rm rad$ and $\theta_{\rm j,max} = 0.7 \ \rm rad $,  $s = -1.63$, $\Estarg = 1.8 \times 10^{51} \ \rm erg$, $\fluxthres = 10^{-7} \ \rm erg \ cm^{-2} \ s^{-1}$,  $\alpha_{pl} = -1.28$, and SFR10. The insets Figures 6(b) and in Figure 6(c) are the same plots as Figures 6(a) and 6(c), respectively, but for the \Swift~data with $\fluxthres = 10^{-8} \ \rm erg \ cm^{-2} \ s^{-1}$ . The solid curves are the complete pre\Swift-Friedman \& Bloom and \Swift-Ryan 2012 redshift distributions. The long dash curves are the calculated distribution assuming a single power-law GRB spectra model with the average low-energy index $\alpha_{pl} = -1.28$. The inset of Figure 6(b) is a plot of a spectral index of the photon flux for a simple power-law model. The data are obtained from the \citet{lie16} paper.  The solid line is the best-fitting line through the data point with an average value of -1.5. The distribution calculations also assume a Hubble constant value of $H_0$ = $72$ km s$^{-1}$ Mpc$^{-1}$. The parameter values in Figures~6(a) and 6(c) also apply to their insets, but not the inset 6(b).}
\label{fig6} 
\finfig
\begfig[t] \hskip-0.25in \epsscale{1.15} \plottwo{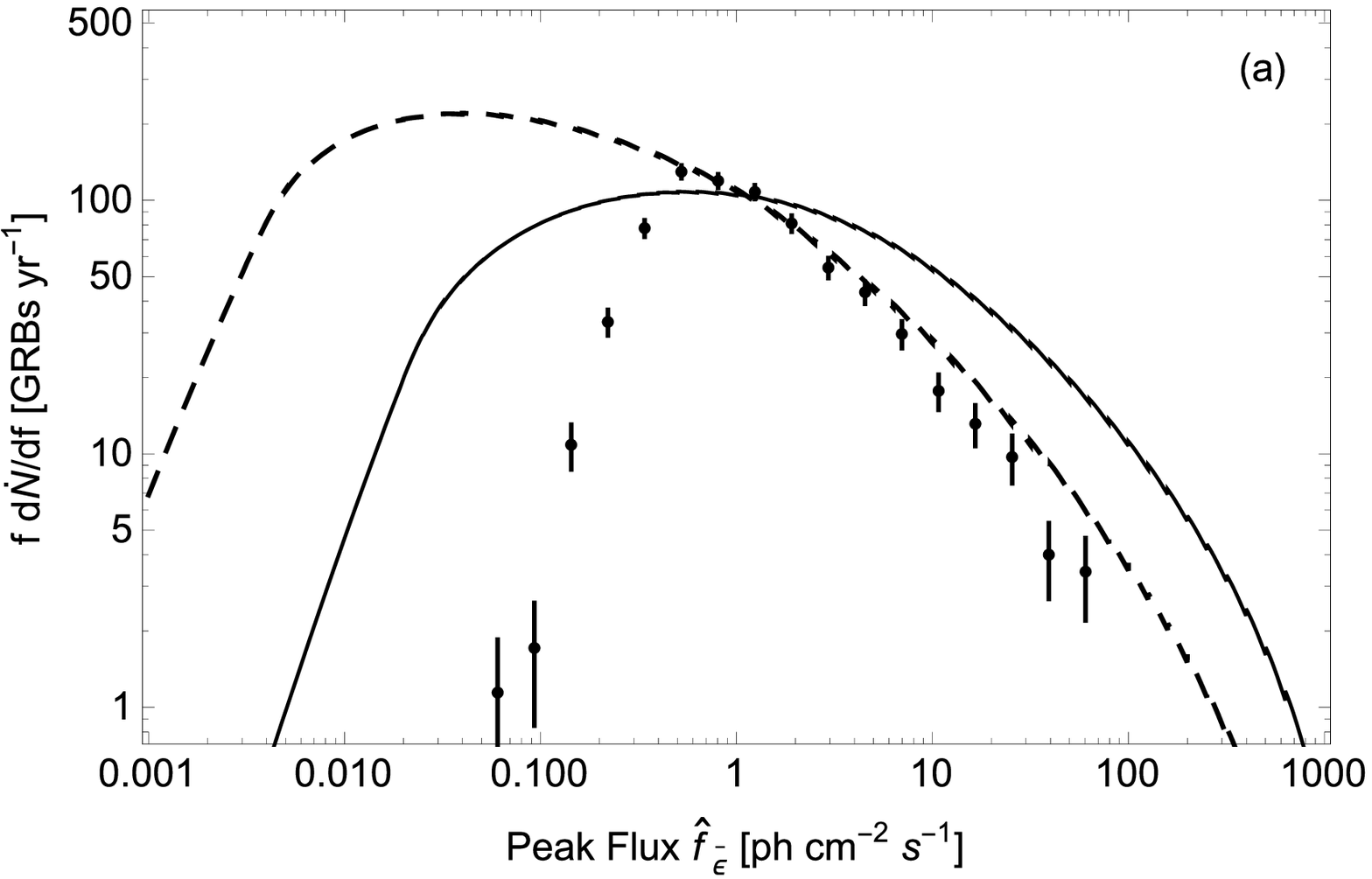}{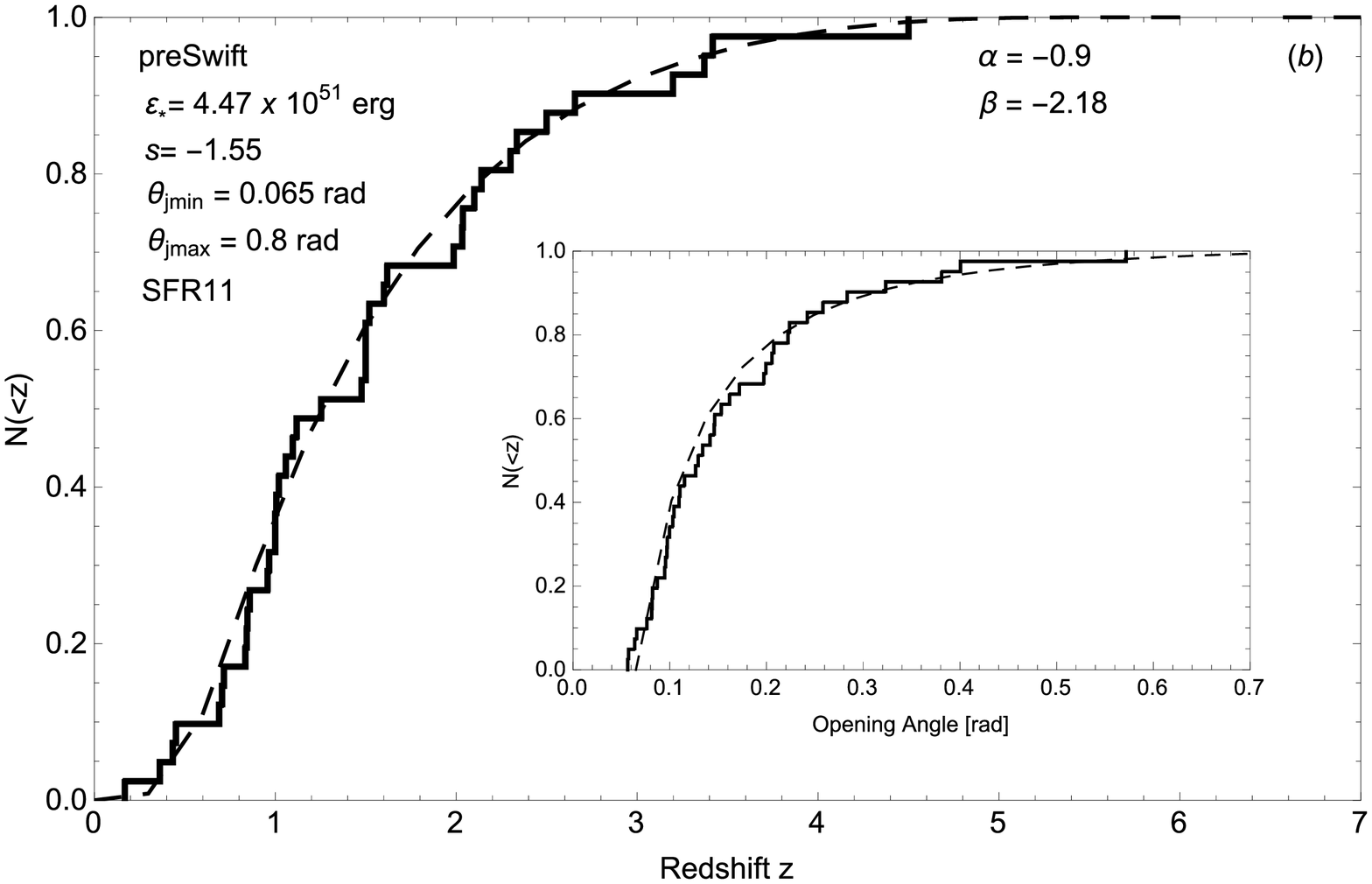}
\caption{\footnotesize (a) The differential size distributions for SFR10 (solid line) and SFR11 (dashed line) with $\Estarg = 2.51 \times 10^{51} \ \rm erg$. The filled circles curve represents the 1024 ms trigger time-scale from the 4B catalogue data, containing 1292 bursts that normalize to 371 LGRBs above a photon number threshold of $0.0633 \ \rm ph \ cm^{-2} \ s^{-1}$, and where the error bars are the statistical error in each bin. Also, the turnover in the data points is due to the detector threshold $0.0633 \rm \ ph \ cm^{-2} \ s^{-1}$ of BATSE. (b) We refit the pre\Swift~redshift distribution and the jet opening angle distribution (inset of Figure 7(a)) using $\Estarg = 4.47 \times 10^{51} \ \rm erg$ and  SFR11, while holding other parameter values the same as shown in Figures~\ref{fig5}(a) and \ref{fig5}(c). The parameter values in Figure~7(b) also apply to its inset.}
\label{fig7} 
\finfig
\begfig[t] \hskip-0.25in \epsscale{1.1} \plottwo{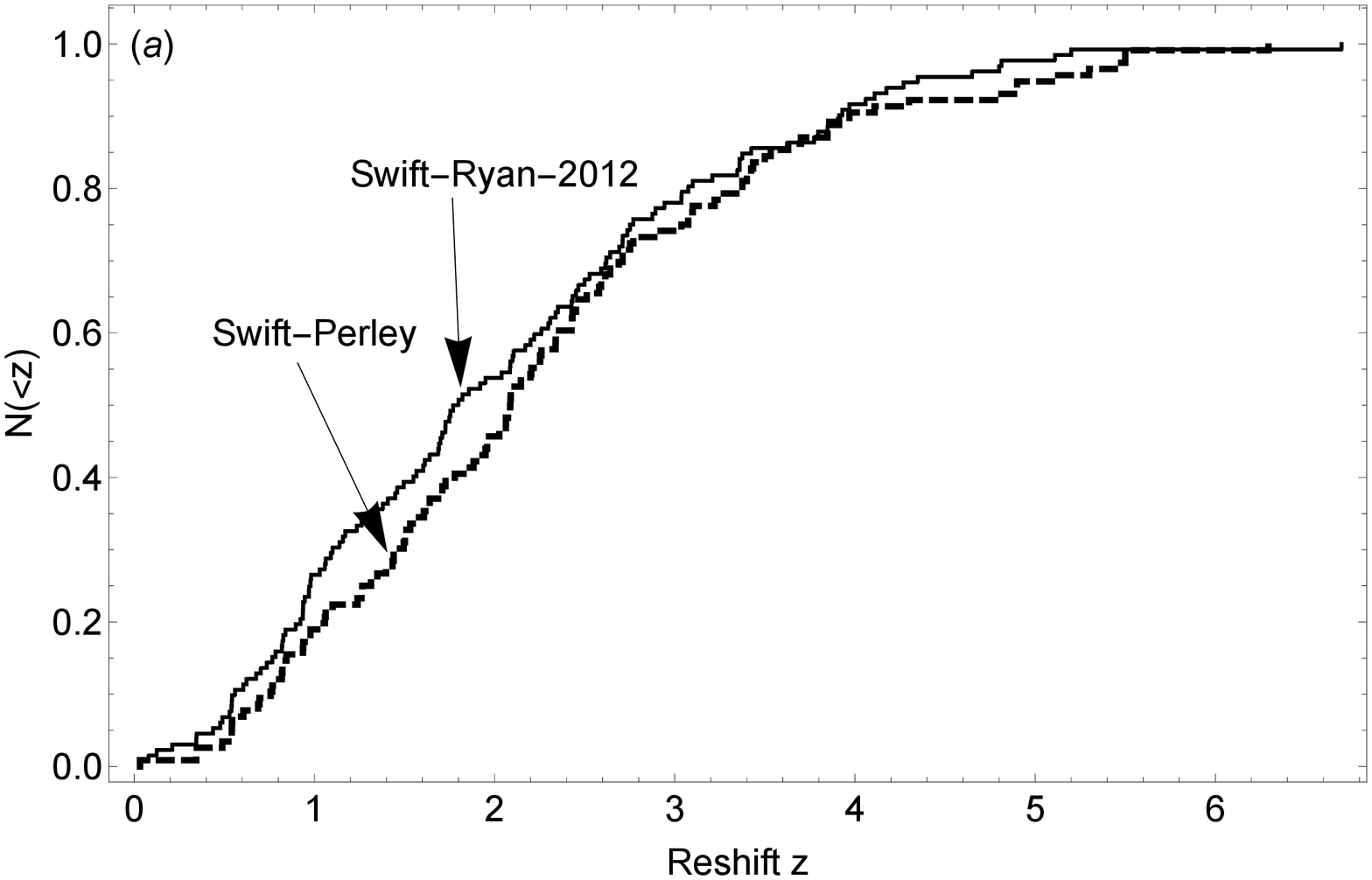}{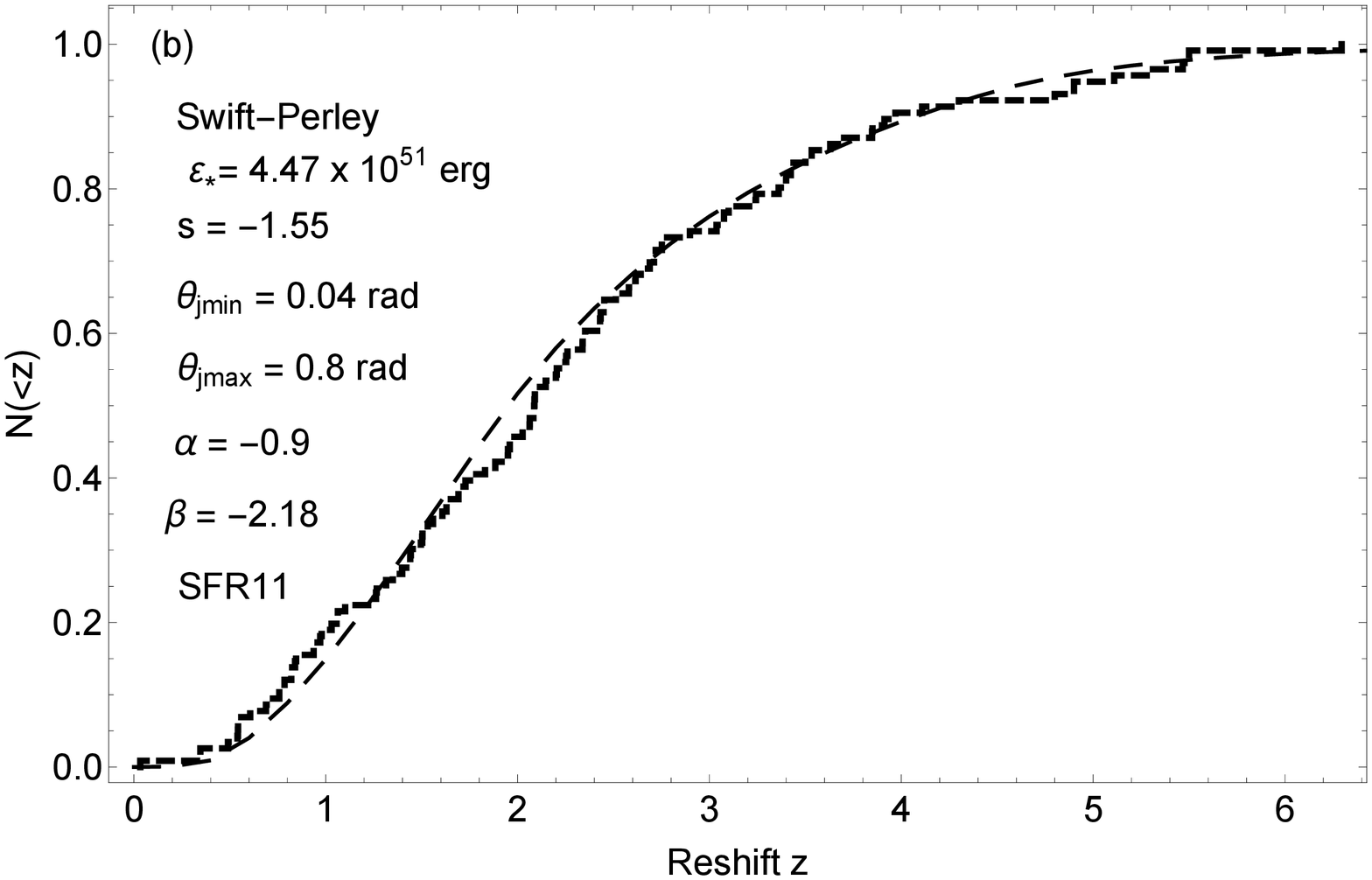}
\caption{\footnotesize ({\it a}) The \Swift-Ryan-2012 and \Swift-Perley samples of the redshift distribution and (b) the fit to redshift of the \Swift-Perley samples using the  pre\Swift~redshift parameters. The long dash and bold curves in Figure~\ref{fig8}(b) are the model and the \Swift-Perley redshifft distributions.}
\label{fig8} 
\finfig
\begfig[t] \hskip-0.25in \epsscale{1.15} \plottwo{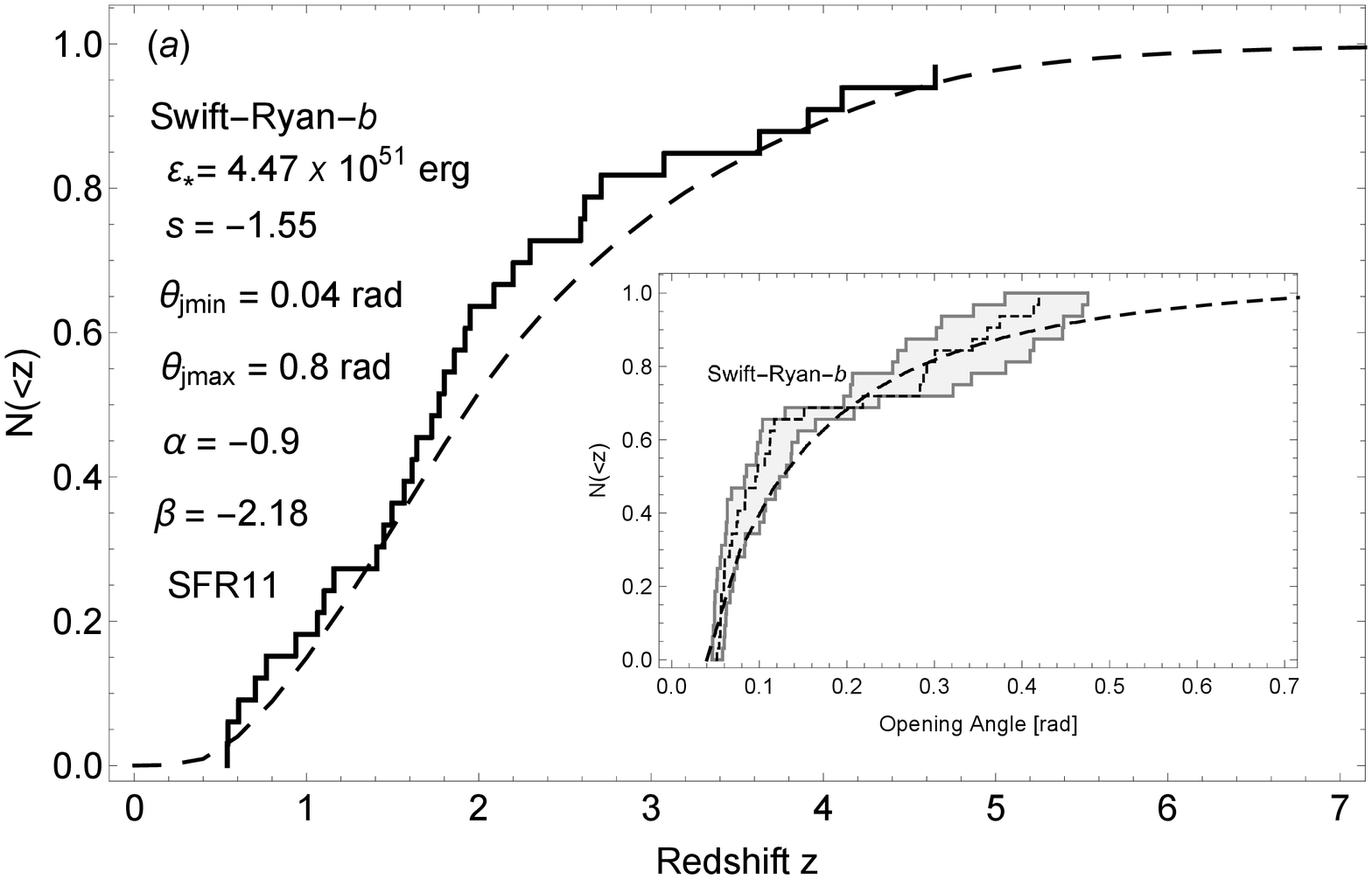}{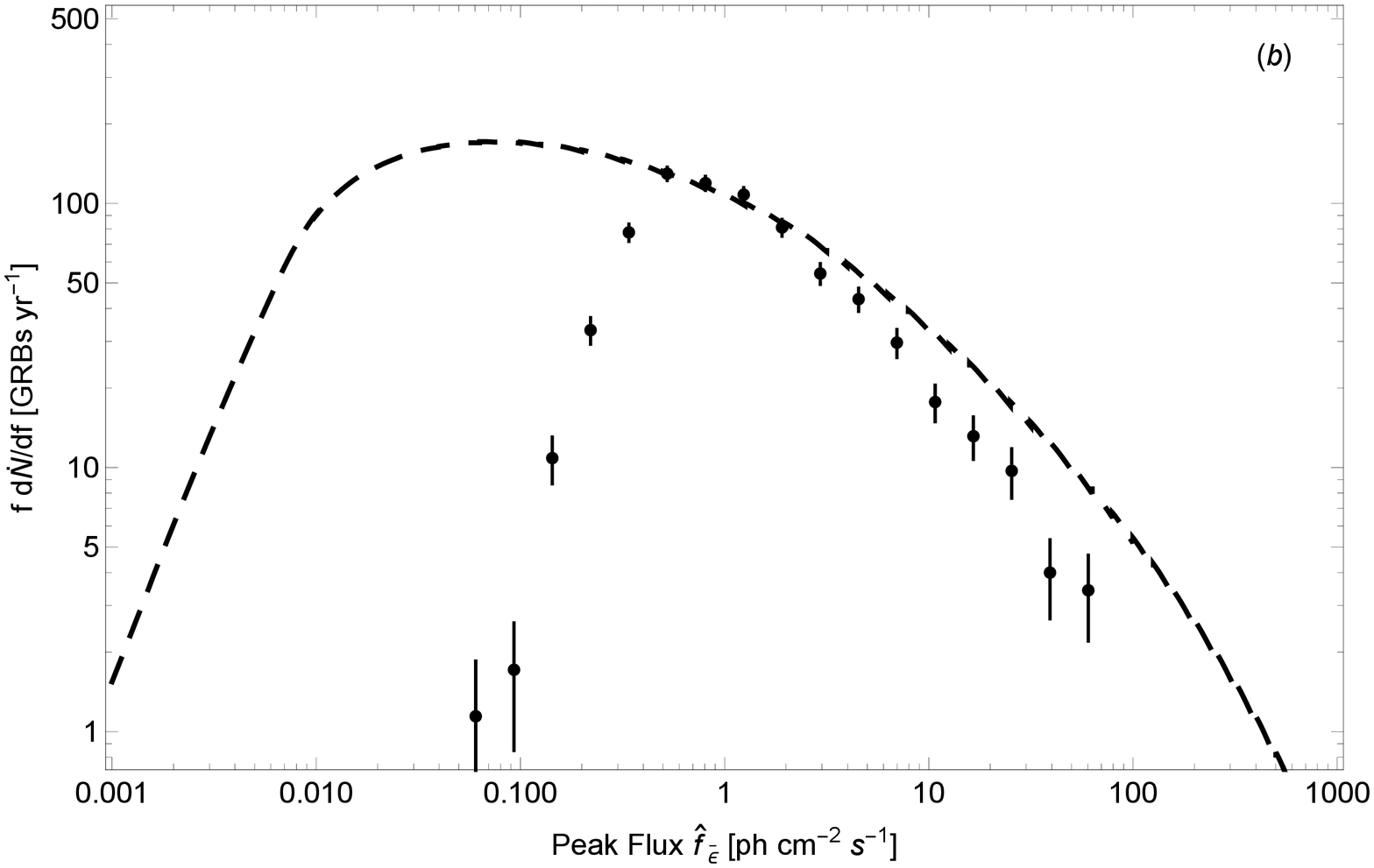}
\caption{\footnotesize (a) The fit to the redshift and opening angle of the \Swift-Ryan-b samples using the  pre\Swift~redshift model parameters in Figure~7(b). In the inset of Figure 9(a), the short dash curve and the shade region in the jet opening angle distribution plot represents the average values and error bars from the Swift-Ryan-b samples. The long dash curves in these Figures are the model distributions. (b) This is the size distribution plot that is similar to Figure~7(a), but for the model parameters in Figure~8(b). The parameter values in Figure~9(a) also apply to its inset.}
\label{fig9} 
\finfig

Recently, \citet{per16} have deduced a large unbiased \Swift~LGRBs sample (\Swift-Perley), as we have discussed in Section~1, by down selecting the \Swift~sample to remove LGRBs that occurred under circumstances that were not optimal for ground-based follow up and isolating a subset for which the afterglow redshift completeness is close to the expected maximum achievable value of 80 per cent. Using their data sample, in Figure~\ref{fig8}(a), we plot the redshift distribution from the \Swift-Ryan 2012 and the \Swift-Perley samples. The plot shows that there is a clear difference in the two samples, however, the Kolmogorov-Smirnov two-sample test indicates that the samples come from the same distribution with values of 0.11 and 0.39 for the $D$-statistic and the $p$-value, respectively. From Figures~\ref{fig8}(b) and \ref{fig9}(a) (including the inset), we plot our model distributions against the \Swift-Perley and the \Swift-Ryan-b samples, respectively, using SFR11 with the same parameter values as for the pre\Swift~sample as indicated in Figure \ref{fig7}(b). The plots show that the observed data are empirically drawn from the proposed model distributions for both the redshift and jet opening angle distributions for the \Swift-Perley and the \Swift-Ryan-b samples. These results indicate that the mean beaming corrected gamma-ray energy released are similar for both pre\Swift~and \Swift~samples, and is a factor of 2 higher than the observed average value (e.g., LD07). The consequence of this is that we would expect to detect more LGRBs at large redshift. Most importantly, the results indicate no excess of LGRB at any given redshift, and our angle distribution implies that the majority of the jet occurs around $\sim 6^\circ$ as shown in the inset of Figure (9a) consistent with \citet{gol16} finding. Using these new parameter values, we recalculate the differential size distribution using SFR11 model and the result is shown in Figure~\ref{fig9}(b). Clearly, the model size distribution is slightly over predicting the differential burst rate above 1.0 ph cm$^{-2}$ s$^{-1}$, which can be slightly improved by reducing the mean beaming gamma-ray energy released with a value between $4.0 \times 10^{51} < \Estarg < 4.47 \times 10^{51} \ \rm erg$ or the jet opening angle power law index between $-1.55 < s < -1.5$. 

Finally, we check for the GRB count rate for \Swift. Our model parameter values, as shown in Figure~\ref{fig8}(b), suggest that GRBs can be detected by \Swift~ to a maximum redshift $z \approx 23$, and that about $10\%$ of \Swift~LGRBs occur at $z > 4.5$, in accord with the data shown in Figure~\ref{fig8}(b). Our model also predicts that about $5\%$ of LGRBs should be detected above $z \geq 5$. If more than $5\%$ of LGRBs with $z > 5$ are detected by the time $\approx 100$ \Swift~LGRBs with measured redshifts are found, then we may conclude  that the LGRB formation rate power-law index $\eta_{_3}$ (see Equation 1) is shallower than our fitted value $\eta_{_3} = -4.1$ above $z \approx 4.5$. By contrast, if a few $z \gtrsim 5$ GRBs have been detected, then this would be in accord with our model and would support the conjecture that the LGRB burst rate follows the SFR11 that is similar to the \citet{hb06} and extended by \citet{li08} SFR history at high redshift. Similarly, our model also indicates that less than $15\%$ of LGRBs should be detected at $z \lesssim 1$ for SFR11. Again, if a few $z \lesssim 1$ GRBS have been detected, then this would also be in accord with our model and would also support the conjecture that the LGRB burst rate follow the SFR11 that is similar to the observed \citet{hb06} SFR history at low redshift with no excess of LGRBs. After examining the GRB samples\footnote{Redshifts were obtained from the catalogues maintained by Dan Perley http://www.astro.caltech.edu/grbox/grbox.php and Jochen Greiner http://www.mpe.mpg.de/$\sim$jcg/grbgen.html.} between 2013 and 2015, we notice that less than $5\%$ of LGRBs per year were detected by Swift above $z > 5$, and about $\sim 10\%$ of LGRBs were detected above $z > 4$, and moreover, only about $6\%$ of LGRBs per year were detected below $z < 1$. These results are consistent with our model predictions, suggesting that our SFR11 model could represent the LGRB density rate.

From the above results, it is clear that our model distributions correlate well with the complete unbiased LGRB \Swift-Perley and the \Swift-Ryan-b  samples indicating no sign of any excess of LGRB at any redshift. While for the complete \Swift-Ryan 2012 sample, our models consistently show that an excess of LGRB is always presence below a redshift of 2, regardless of how we modify our model parameter values to fit both the pre\Swift~and \Swift~distributions. Interestingly, \citet{pes16} reach similar conclusion when incomplete sample was used. This suggests that there could be a biased in the \Swift-Ryan-2012 sample relating to the afterglow selection effect based on the \Swift-Perley sample; or that the \Swift-Ryan 2012 sample needs to be constrained using the ``well-fitted'' beaming opening angle. These results could be checked, but they are beyond the scope of this paper.

\section{CONCLUSIONS}
The question whether LGRBs follow SFR is now getting closer.  Le \& Dermer developed a physical model to understand the differences between the redshift and jet opening angle distributions of \Swift~and pre\Swift~GRBs, taking into account the different detector triggering properties. Their GRB model is based on the uniform jet model and a flat GRB spectrum with an assumption that the LGRB density rate is proportional to the measured SFR. They showed that a good fit was only possible by providing a positive evolution of the SFR history of GRBs to high redshifts. However, it remains unclear whether this excess at high redshift is due to luminosity evolution or the cosmic evolution of the GRB rate. Le \& Mehta revisited the work done by LD07 and performed a timely study of the rate density of GRBs with an assumed broken power-law GRB spectrum. Utilizing more than 100 LGRBs in the \Swift~sample that includes both the observed estimated redshifts and jet opening angles, they obtained a GRB burst rate functional form that gives acceptable fits to the pre\Swift~and \Swift~redshift and jet opening angle distributions with a GRB density rate, SFR9, that is similar to the \citet{hb06} observed star formation history (SFR7) and as extended by \citet{li08}. However, the LT17 redshift model distribution indicates an excess of LGRBs at low redshift below $z \sim 2$ in the \Swift~sample, consistent with \citet{pkk15}, \citet{yu15}, and \citet{laj19}. However, the reason for this excess is either unclear, incomplete sample size, or that GRBs formation rate does not trace  SFR at low redshift less than $z \leq 1$. As a result, in this paper, our motivations are to determine  (i) how does the observed Hubble constant parameter affect the outcome of the calculated redshift and jet opening angle distributions? (ii) How does the beaming-corrected gamma-ray energy released relate to redshift? (iii) How do the low- and high-energy indices ($\alpha, \beta$) of the Band function relate to redshift?

Each point has been addressed in our paper and we conclude that: (i) the Hubble constant is insensitive to the calculated redshift distribution, but might not be for the directional event rate per unit redshift distribution. We plan to explore this possibility in future paper. (ii) The mean beaming corrected gamma-ray energy released is similar between the \Swift~and pre\Swift~samples or about a factor of two higher for the pre\Swift~data. However, if the beaming corrected energy released is similar for both instruments, then the \Swift~jet opening angle should be smaller at higher redshift. Interestingly, our model provides consistent result with the observed estimated jet opening angle for both the \Swift-Ryan-b and pre\Swift~samples. Our best value for the mean beaming corrected gamma-ray energy released is $\Estarg \sim 4.47 \times 10^{51} \ \rm erg$ using \Swift-Perley sample, which is about a factor 2 higher than the observed average value (e.g., LD07). Moreover, our results also agree with the observed minimum jet opening angle of $\theta_{\rm j, min} \approx 0.065$ and $\approx 0.4$ rad for the pre\Swift~and \Swift-Ryan-b jet opening angle samples, respectively, with the maximum jet opening angle of $\theta_{\rm j, max} \approx 0.8$ rad and the $s = -1.55$ for the jet opening angle power-law index. (iii) We find that the low- and high-energy indices ($\alpha, \beta$) of the Band function are independent of redshift and the best values are about $\alpha \approx - 0.9$ and $\beta \approx -2.18$ and these values are similar to the observed values. 

Using SFR11, with $\eta_{_1} = 5.5, \eta_{_2} = 0.38, \eta_{_3} = -4.1, z_1 = 0.5$, and $z_2 = 4.5$ for the GRB formation rate in Equation(\ref{eq1}), our model gives a good fit to the pre\Swift, \Swift-Ryan, \Swift-Ryan-b, and \Swift-Perley redshift and jet opening angle samples. However, we find that our model is more consistent with the pre\Swift, \Swift-Ryan-b, and \Swift-Perley samples with a single value of the mean beaming-corrected gamma-ray energy released of $\Estarg \sim 4.47 \times 10^{51} \ \rm erg$ or lower. Furthermore, this model also predicts about $5\%$ of the \Swift-LGRBs should be detected above $z \geq 5$ and less than $15\%$ of the \Swift-LGRBs should be detected at $z \lesssim 1$. Examining the \Swift~LGRB sample between the years 2013 and 2015, the  data show consistent results against our model predictions. This suggests that the values $\eta_{_1} = 5.5, \eta_{_2} = 0.38$, $\eta_{_3} = -4.1$, and the redshift breaking points at $z_1 = 0.5$ and $z_2 = 4.5$ in our SFR11 model provide an LGRB formation rate that is consistent with observation. Additionally, the values at $z_1$ and $z_2$ seem to correlate well between the model and the LGRB observed data for the directional event rate per unit redshift per solid angle calculation, which we have touched on in Figure~\ref{fig2}. The results of this correlation will be presented in future paper. And more importantly, SFR11 produces a general behavior of the observed SFR that agrees with the conjecture that LGRB rate follows SFR. However, the GRB formation rate (SFR11) is higher than the observed SFR (SFR7) above a redshift of $z > 3$, and this has been suggested to be related to metallicity \citep[e.g.,][]{fru06,gr13}. For example, \citet{fru06} and \citet{gr13} have suggested that LGRBs are found more in regions of low metallicity (at high redshift) and this seems to be consistent with our GRB formation rate SFR11 model. We plan to explore metallicity evolution in future work. Moreover, in the region of redshift below $z < 2$, our SFR11 follows the \citet{hb06} observed SFR (SFR7), and no excess of LGRBs is implied based on the \Swift-Perley sample. In the next few years if more LGRBs have been detected at redshift less than $\sim 2$ or $z > 3$, then the values of $\eta_{_1}$, would be more positive, and $\eta_{_2}$ and $\eta_{_3}$ would be more negative, assuming all the other physical parameters ($H_{\rm 0}, \alpha, \beta, \theta_{\rm j, min},  \theta_{\rm j, max}, \Estarg$, and $\fluxthres$) remain the same.

Additionally, our analyses indicate that none of the above physical parameters resolved the excess problem from the \Swift-Ryan-2012 sample, but our model is well fitted with the \Swift-Ryan-b sample with no excess problem, suggesting that the \Swift-Ryan-2012 sample is biased with sample selection effect. Using the \Swift-Perley LGRB sample (Swift Gamma-Ray Burst Host Galaxy Legacy Survey) and applying the same physical parameter values that provides the best fit to the pre\Swift~and \Swift-Ryan-2012 samples, our model provides consistent results with this data set and indicate no excess of LGRBs at any redshift. Since our model shows strong correlations with the \Swift-Perley sample, we conclude that there could be a biased in the \Swift-Ryan sample relating to the afterglow selection effect. We plan to check if this is the case in future work.

\acknowledgements TL wishes to acknowledge Peter Veres for discussions and comments, and the anonymous referee for many useful comments and suggestions that greatly improved the paper.

\end{document}